\documentclass{emulateapj-rtx4}

\usepackage{rotating}
\usepackage{epsf}
\usepackage [english]{babel}
\usepackage [autostyle, english = american]{csquotes}
\MakeOuterQuote{"}

\begin{document}
ACCEPTED FOR PUBLICATION IN ApJ

\title{X-ray Emission from the Taffy (VV254) Galaxies and Bridge}

\author{P. N. Appleton\altaffilmark{1}, L. Lanz\altaffilmark{1}, T. Bitsakis\altaffilmark{2}, J. Wang\altaffilmark{3}, B. W. Peterson\altaffilmark{4}, U. Lisenfeld\altaffilmark{5}, K. Alatalo\altaffilmark{1,6,$\dagger$},
 P. Guillard\altaffilmark{7,8}, F. Boulanger\altaffilmark{9}, M. Cluver\altaffilmark{10}, Y. Gao\altaffilmark{11}, G. Helou\altaffilmark{1},  P. Ogle\altaffilmark{1} and C. Struck\altaffilmark{12}}
\altaffiltext{1}{Infrared Processing and Analysis Center, California Institute of Technology, MC100-22, Pasadena, California 91125, USA; apple@ipac.caltech.edu}
\altaffiltext{2}{Instituto de Astronomia, Universidad Nacional Autonoma de Mexico, A.P. 70-264 D.F., Mexico}
\altaffiltext{3}{Department of Astronomy and Institute of Theoretical Physics and Astrophysics, Xiamen University, Xiamen, Fujian 361005, China}
\altaffiltext{4}{University of Wisconsin-Barron County, Rice Lake, WI 54868, USA}
\altaffiltext{5}{Dept. Fisica Teorica y del Cosmos, University of Granada, Edifica Mecenas, Granada, Spain}
\altaffiltext{6}{Observatories of the Carnegie Institution of Washington, 813 Santa Barbara Street, Pasadena, CA 91101, USA}
\altaffiltext{7}{Sorbonne Universites, UPMC Univ. Paris 6 et CNRS, UMR 7095, France}
\altaffiltext{9}{Institute d'Astrophysique de Paris, 98 bis Boulevard Arago - 75014 Paris, France}
\altaffiltext{8}{Institut d'Astrophysique Spatiale, Universite Paris Sud 11, Orsay, France}
\altaffiltext{10}{University of the Western Cape, Cape Town, South Africa}
\altaffiltext{11}{Purple Mountain Observatory, Nanjing, China}
\altaffiltext{12}{Iowa State University, Ames, IA 50011}

\altaffiltext{$\dagger$}{NASA Hubble fellow}

\begin{abstract}
We present the first X-ray observations of the Taffy galaxies (UGC 12914/5) with the{\it~Chandra} observatory,  and detect soft X-ray emission in the region of the gas-rich, radio-continuum-emitting Taffy bridge. The results are compared to{\it~Herschel} observations of dust and diffuse [CII] line-emitting gas. The diffuse component of the Taffy bridge has an X-ray luminosity of L$_{X(0.5-8keV)}$ ~=~5.4 $\times$ 10$^{39}$~erg~s$^{-1}$, which accounts for 19$\%$ of the luminosity of the sum for the two galaxies. The total mass in hot gas is (0.8--1.3)~$\times$~10$^8$~M$_{\odot}$, which is approximately 1$\%$ of the total (HI~+~H$_2$) gas mass in the bridge, and $\sim$11$\%$ of the mass of warm molecular hydrogen discovered by{\it~Spitzer}.  The soft X-ray and dense CO-emitting gas in the bridge have offset distributions, with the X-rays peaking along the north-western side of the bridge in the region where stronger far-IR dust and diffuse [CII] gas is observed by{\it~Herschel}. We detect nine Ultra Luminous  X-ray sources (ULXs) in the system,  the brightest of which is found in the bridge, associated with an extragalactic HII region. We suggest that the X-ray--emitting gas has been shocked--heated to high temperatures and "splashed" into the bridge by the collision. The large amount of gas stripped from the galaxies into the bridge and its very long gas depletion timescale ($>$10 Gyr) may explain why this system, unlike most major mergers, is  not a powerful IR emitter.  

\end{abstract}


\section{Introduction}

Galaxy collisions and interactions play an important role in the local universe in driving powerful star formation activity in galaxies, and the generation of powerful emission in the infrared \citep{jos85,soi87,soi91}. Multi-wavelength studies of local galaxy populations have shown that nearly all Ultraluminous Infrared Galaxies (ULIRGS with L$_{IR}$ $>$ 10$^{12}$ L$_{\odot}$) involve mergers of gas-rich galaxy pairs \citep{arm87,san88a,san88b,san96}. Similarly, nearby Luminous IR Galaxies \citep[LIRGS with L$_{IR}$ $>$ 10$^{11}$ L$_{\odot}$, see][]{elb02}, often involve strongly interacting and sometimes merging systems. Numerical models \citep{bar96} have emphasized the importance of strong tidal torques, which can act to  drive gas into the centers of galaxies, initiating nuclear enhancements in star formation. Despite two decades of study of these galaxies, there are still many uncertainties about how to realistically model the complex astrophysical environment of a powerful major merger \citep{hop09,hop14,hay14}. Many questions remain.  For example, what is the the dependence of the star formation triggering on the physical conditions in the gas, especially the effects of feedback from young stars, active galactic nuclei (AGN) winds,  or other processes \citep[see][]{hop14}.  Do all deeply penetrating gas-rich mergers inevitably lead to powerful star formation and luminous infrared activity? 

There are examples of major mergers where the gas has been disturbed
significantly, but is still in a  pre-(U)LIRG phase. The `Antennae' system \citep[NGC 4038/39;][]{whi95,mir98,wil03} is only just classified as a LIRG, and yet clearly contains large amounts of gas
in the overlap region which have not fully developed their star forming potential. Clues to this are
seen in the recent high-resolution study of CO(3--2) with ALMA by \citet{whi14}, where
peculiar velocities and compact linear  filaments of CO gas are seen: likely the result of turbulence. 
In the current paper, we discuss another possible example of a galaxy pair, the Taffy (total L$_{FIR}$ = 4.5 $\times$ 10$^{10}$ L$_{\odot}$),  that has recently experienced a head-on collision but has yet to display strong global star formation enhancements. 

The Taffy galaxies, UGC 12914 and UGC 12915 (VV254, KPG603)  were discovered to be connected by a bridge of radio continuum emission (Figure 1a) and neutral hydrogen gas \citep{con93}. It was suggested that the two galaxies have recently interpenetrated, creating a common magnetic field between them as they moved apart after the collision. In this picture, cosmic rays (CRs) generated within the galaxies, would flow into the common halo generating radio synchrotron radiation.  \citet{con93} showed that the logarithmic ratio of the far-infrared to radio continuum flux $q$ is 1.94, unusually low for normal galaxies ($<q>$ = 2.3).  

\citet{lis10} suggested that {\it in situ} shock acceleration of cosmic rays in the bridge can explain the strength of the radio emission and radio spectral index. In this view, the galaxies do not need to be the source of CR electrons, because the CRs can be accelerated within the highly turbulent conditions in the bridge \citep{con93,bra03,zhu07,gao03a}. 

Further evidence that shocks may play an important role in the material in the bridge was presented by \citet{pet12}, who showed that the bridge contains $>$ 4.5 $\times$ 10$^{8}$ M$_{\odot}$ of warm (150 $<$ T $<$ 175 K) molecular hydrogen.  The excitation of this warm intergalactic gas was found to have a strong heating contribution from shocks and turbulence. The detection of broad CO (1--0) lines in the bridge \citep{gao03a} is consistent with turbulent gas and kinetic energy dissipation in this colder component.  This could provide a significant fraction of the heating needed to maintain the warm H$_2$ for the $\sim$25 Myrs since the collision \citep{con93}.  While \citet{pet12} were able to demonstrate that the warm H$_2$ emission was likely the result of turbulent or shock--heating of the bridge, they could not rule out heating of the molecular gas by X-rays, since, at that time, no observations of this kind had been made.
 
One of the consequences of a head-on collision between the two galaxies is the possibility of the sweeping of gas out of the galaxies. The large amounts of gas found in the bridge \citep[$\sim$7 $\times$ 10$^{9}$ M$_{\odot}$,][]{zhu07,vol12,pet12} has to survive the violence of the collision \citep{bra03}. Nevertheless, in the gas and stellar dynamical modeling of the Taffy collision by \citet{vol12}, it was shown that momentum transfer between the gas clouds in the galaxies during the collision was sufficient to transport large quantities of gas into the bridge.  The model also provided a possible explanation for the spatial offset in the distribution of the dense molecular gas, and the diffuse HI in the bridge \citep{gao03a}. This was explained as a consequence of the two galaxies colliding slightly off-center, leading to an asymmetry in the density structure of the resulting bridge. The models did not include the thermal balance of the different phases of the interstellar medium (ISM) in the galaxies, which is especially relevant to X-ray observations. 
  
Hot X-ray emitting gas can extend outside galaxy disks through both internal and external processes. In starburst galaxies, supernova explosions can create hot bubbles of X-ray emitting gas that can break-out of their host molecular disks, and push gas into the halo \citep{wat84,fab84,hec90,hec02}.  Large scale shocks can also heat gas to X-ray temperatures in the intergalactic diffuse environment. In the 40 kpc-long filament in the Stephan's Quintet galaxy group, X-ray emission was detected in association with a giant shock structure \citep{tri03,tri05,osu09}.  The filament also contains copious amounts of warm and cold molecular gas \citep{app06,clu10,gui12}. The emission is consistent with the model of \citet{gui09} in which a large-scale shock is propagated into a multi-phase medium. In this model, the low density regions of this medium become heated to X-ray temperatures, whereas the higher density clumps collapse and can survive the initial disturbance and radiate strongly in the mid-IR H$_2$ lines.  Shocked regions can also radiate strongly in diffuse [CII] emission, as observed \citep{app13}. Similarly powerful mid-IR H$_2$, [CII], and [OI] emission have been seen in the Taffy system (Peterson \& Appleton, in Preparation), providing further motivation for the current study of the Taffy system in X-rays.  

In this paper, we detect faint soft diffuse X-rays from the Taffy galaxy disks and bridge region, as well as from a number of luminous discrete sources.  We explore various explanations for the observed X-ray emission, favoring shock--heating as the most likely explanation for the diffuse emission. We assume a distance to the galaxies of 62 Mpc based on the heliocentric velocities for UGC 12914 and UGC 12915 of $4371\pm8$ and $4336\pm7$ km s$^{-1}$ respectively, and a Hubble constant of 70 km s$^{-1}$Mpc$^{-1}$.

\section{Observations}

\subsection{Chandra Observations}
The Taffy galaxies were observed for 39.5\,ks on 2013 February 3 (ObsID 14898; PI P. Appleton) with the back-illuminated CCD chip, S3, of the {\it Chandra} Advanced CCD Imaging Spectrometer \citep[ACIS;][]{wei00}. We processed the observation using {\sc ciao} version 4.6 (Calibration Database 4.6.4) to create a level--2 events file, following the software threads from the {\it Chandra} X-ray Center (CXC)\footnote{\url{http://cxc.harvard.edu/ciao}}. 

Most of the images, covering a 4.1 x 4.1 arcminute$^2$ field with a pixel scale of 0.492 arcsec pixel$^{-1}$ are shown for the $0.5$--$8$\,keV range, but we also examined smaller energy ranges, including soft ($0.5$--$1$\,keV and $0.5$--$2$\,keV), medium ($1$--$2$\,keV), and hard ($2$--$8$\,keV) bands.  We created smoothed images in two ways. First, we smoothed with a simple $1\farcs5$ Gaussian kernel with the SAOImager ds9 \citep{joy03}. Second, we used the {\sc ciao} task {\sc dmimgadapt} to adaptively smooth images. {\sc dmimgadapt} smooths each pixel on the scales necessary to reach the desired count threshold (selected to be 10 counts) under its convolution kernel or the maximum kernel size, which we set to 5$''$. 

We also created images of the diffuse emission by removing point sources, dividing by the exposure map, and then adaptively smoothing the cleaned image (Section 3.2). Point source identification was done using the {\sc ciao} task {\sc wavdetect} \citep{freeman02}, a Mexican-Hat wavelet source detection program. These sources, except for the plausible centers of the galaxies, were then cleaned using the task {\sc dmfilth}, which replaces source pixel values with values determined by interpolating the surrounding background. 

We used the {\sc specextract} task to extract X-ray spectra in the 0.3--8.0\,keV range for several extraction apertures in the Taffy system, including the two galaxies and the bridge, after excluding the point sources. A background spectrum was simultaneously extracted in the same chip in a sourceless region, and automatically scaled based on the ratio of the source-to-background areas. We filtered the events based on the energy range and then grouped it to a minimum of 20 counts per bin prior to modeling the spectrum. For the two bridge sub-regions as well as the extragalactic HII complex (see Section 3), we estimate the flux and luminosities based on the count rates, assuming the model parameters from the total bridge. 

\subsection{Spitzer and Herschel Observations}

The {\it Spitzer} 24$\mu$m observations shown in this paper were obtained from the Spitzer Heritage Archive (PI: J. R. Houck; Observid 21) and had been processed through the S18 SSC pipeline. A constant background was removed from the image before extracting fluxes from the extended regions. 

The {\it Herschel} photometry data at 70, 100 and 160$\mu$m\footnote{{\it Herschel} is an ESA space observatory with science instruments provided by European-led Principal Investigator consortia and with important participation from NASA.} was de-archived from the Herschel Science Archive (PI E. Sturm, Proposal name KPGT$\_$esturm$\_$1, Obsids 1342212798/9) as Level 1 data, which are fully calibrated frames. These scan and cross-scan data were then further processed through the mapping package SCANAMORPHOS (Roussel 2013), kindly made available to us by the author.   Additionally, we obtained {\it Herschel} [CII]157.7$\mu$m line data as part of our OT1 program PACS spectroscopy program (PI: P. Appleton, proposal OT1$\_$pappleto$\_$1, Obsid 1342238415). These data were processed on a large virtual machine through a version of HIPE 13 \citep{ott10} available at the NASA Herschel Science Center. The integrated line map shown here was created in HIPE after first removing baselines from all the individual spaxels in the image, and then integrating over the range of velocities in which signal was detected.  Full details of the reduction of the [CII] data will be provided in a separate paper devoted to the PACS spectroscopy of the Taffy system (Peterson \& Appleton, in Preparation).     

\section{Results}

\subsection{X-ray morphology}

We show in Figure 1b and 1c two different representations of the observations. In Figure 1b, we show the compact X-ray structure present in the 0.5--8 keV image of these data. We detect nine bright point-like sources at high significance above background, and numerous fainter ones.   An additional extended source, is coincident with the peak of the radio continuum emission in the nucleus of the south-western galaxy UGC 12914 (and 2MASS K-band emission-not shown).  In contrast, the north-eastern galaxy, UGC 12915, contains several point-like X-ray sources, but none of them seem to be  directly associated with the galaxy's radio continuum and near-IR light peak.  Fig. 1c shows clearly the detection of extended emission in the bridge region, but it is somewhat offset from the radio continuum bridge to the north-west. 

To quantitatively describe the X-ray properties of the various features seen in the X-ray images, we define in Figure 2a several extended polygon-shaped regions, as well as identifying the point sources. For the extended regions, polygons define both galaxies and the bridge, which is divided into two parts\footnote{ The reason for this specific division of the bridge relates to the separation between the mainly denser molecular gas and the diffuse HI. This is discussed further in Section 4.}. 
The X-ray properties of the galaxies are given in Table 1, along with coordinates, aperture description, net counts (after taking into account the background model), and total X-ray flux (0.5--8 keV) and luminosity. The luminosities are derived from model fits to the X-ray spectra (Section 3.3).
The X-ray luminosities (0.5--8 keV) of UGC 12914 and UCG 12915 (excluding the point sources) are 1.66 and 1.15 $\times$10$^{40}$ erg s$^{-1}$ respectively. If we add the contribution of the compact sources within their disks (see next section), UGC 12914 and UGC 12915 have total luminosities of 3.2 and 2.2 $\times$10$^{40}$ erg s$^{-1}$ respectively. Therefore, the compact sources contribute significantly to the total X-ray luminosities of the galaxies.  

In Table \ref{soft}, we present the observed soft (0.5--2 keV) X-ray fluxes for the extended polygon regions. We measure a total X-ray luminosity over the the full bridge of  L$_{X(0.5-2keV)}$ = 4.7 $\times$ 10$^{39}$ ergs s$^{-1}$.  Most of the soft X-ray emission from the bridge comes from region Bridge-1, with a
surface X-ray brightness of 5.7 $\times$ 10$^{-18}$ erg s$^{-1}$ cm$^{-2}$ arcsec$^{-2}$.  This is only a little less that the average surface brightness of the soft X-ray emission from the two galaxies.  The region called "HII", which we consider separately from the bridge (Figure 2a), is dominated by the compact source B1, but may also include some extended emission.

\subsection{The Compact Sources}

Figure 2a and Figure 2b shows two kinds of labeled point sources. Those with yellow circles E1-E5 (in UGC 12915), W1-W3 (in UGC 12914), and B1 in the bridge, are all detected at a 3-sigma level of significance. The level of detection significance is based on Poisson noise estimates, taking into account the estimated (generally low) background counts detected in addition to the sources.  These sources, which are strongly detected, are listed individually in Table 1. We further identify 15 sources (labelled as red circles in Figure 3) which are detected at a lower level of significance (2-3$\sigma$). These sources have net counts in the  1.5-2 arcsecs extraction apertures of between 4.5 and 10.  Formally they represent significant detections, but their X-ray luminosities are poorly determined because of Poisson noise. We list the properties of these sources in Table \ref{compact}. 
How many of the fainter sources would we expect in the area of the bridge ($\sim$1 arcmin$^{-2}$) by chance? Based on an exploration of larger adjacent fields on the chip, and by simulating an additional component representing the added (randomly distributed) count rate from the diffuse Taffy-bridge emission, we  would expect  0.3 sources arcmin$^{-2}$ with a count rate $>$ 4 in a 1.5 arcsec extraction aperture.  This number is consistent with the expected number of AGN (0.5 sources arcmin$^{-2}$) based on a detection threshold of close to 1.5 $\times$ 10$^{-15}$ erg s$^{-1}$ cm$^{-2}$ \citep{har03}. Considering the larger area of the whole Taffy system, we conclude that no more than one of the fainter sources is a chance association: most appear to be real detections.

To investigate how their properties might influence our estimate of the extended soft X-ray emission from the bridge, we identified them by eye\footnote{Only two of them were detected formally by WAVDETECT (Freeman et al. 2002). This may not be surprising, since WAVDETECT may not detect all sources close to the detection threshold.} and flagged them for removal in the smoothed images.  As can be seen by comparing Figure 2c with Figure 2d, the effects of the removal of these additional sources is small, reducing the net flux in the soft band by only an additional 20$\%$. In what follows, we present results based on the extended emission with only the more reliable 3$\sigma$ sources removed. 

We estimate the flux of each 3$\sigma$-detected source assuming a power law with a photon index of $\Gamma$=1.7 \citep{swa04,sor11}. Their luminosities (Table 1) put them in the range of Ultra-luminous X-ray sources \citep[ULXs; commonly associated with star forming galaxies][]{swa11,lua15}. Such sources are usually assumed to be massive accreting black-holes (M$_{BH}$ $\leq$ 20 M$_{\odot}$) associated with young  star clusters \citep[see][]{zez07,min12a}. This is consistent with the majority of the compact sources being associated with the disks of both galaxies, as shown in Figure 3. Several are concentrated around the inner region of UGC 12915, and some of the fainter sources seem loosely associated with the bright spiral arm in UGC 12914, seen in visible light.  The numbers of brighter sources are also consistent with the average number of ULX's as a function of star formation rate (SFR). \citet{swa11} estimate that for gas-rich galaxies, there is, on average, 0.5 ULXs for every 1 M$_{\odot}$ yr$^{-1}$ of SFR. Based on the SFRs for the two galaxies discussed in Section 5.1, we estimate that UGC 12914 and UGC 12915 should contain 2 and 5 ULXs (L$_{X(0.3-8keV)}$ $>$ 10$^{39}$ erg s$^{-1}$), which is close to the actual number of 3$\sigma$ sources found in the two disks. Unlike \citet{smi12}, who studied a sample of interacting galaxies, we do not see a significant enhancement in ULXs compared with normal spirals 
in this system.  
 
The brightest ULX source (B1) lies outside the disk, to the south-west of UGC 12915, and has an X-ray luminosity of L$_{X(0.5-8keV)}$ = 3.0 $\times$10$^{39}$ erg s$^{-1}$.  This bridge source is associated with a massive molecular rich cloud-complex \citep[see][]{gao03a}, and is the only region outside the galaxies detected in H--$\alpha$ and Pa--$\alpha$ images \citep{bus87,kom12}.  We estimate that this extragalactic HII region complex (labelled as "HII" in Figure 2a) has a  SFR $\sim$ 0.25 M$_{\odot}$ yr$^{-1}$ (Section 5.1 of this paper) based on its IR emission. It also lies close to a peak in the radio emission in the bridge seen in Figure 1a. As such, the discovery of a single ULX in this complex is interesting, as it lies outside either of the two galaxy disks and lies near the densest part of the molecular bridge \citep[][and Braine et al. in preparation]{gao03a}.  The failure to detect the source in hard X-rays probably also rules out a chance association with a background AGN. Although no age estimate is available for the source (it is highly obscured in visible and UV light), the escape of emission of H--$\alpha$ light suggests a young stellar population is present. If this "extragalactic HII region" was formed by a compression of gas in the bridge, it would only have had 10--20 Myrs to form. However, this is feasible, since  Berghea et al. (2013) have shown that the ages of typical massive star clusters associated with ULXs are typically 10 Myrs old. A more detailed study of the correlation between the position of Taffy B1 and the host molecular cloud and associated star-formation would be worthwhile.

In addition to B1, several fainter sources lie in the bridge (B2-5). Although we cannot completely rule of a chance association with a background source, the concentration of these fainter sources is interesting. As Figure 3 shows, the sources fall in the region where faint dusty filaments are seen in the optical image--suggesting heavy obscuration, and they fall well within the main region of the molecular bridge (Section 4). It is therefore possible that these sources represent fainter versions of B1, and are associated with obscured star formation within the bridge. Unlike B1, there is no clear association of these sources with optical counterparts.


The slightly extended source "W$_{nuc}$" lies in the nucleus of UGC 12914, and contains a significant fraction of the X-ray luminosity of the whole galaxy. This source is the only source detected in the 2--8 keV band, with a hard X-ray luminosity L$_{X(2-10keV)}$ = 1.1$\pm$0.2 $\times$ 10$^{40}$ erg s$^{-1}$ (luminosity calculation assumes an X-ray  power-law of slope $\Gamma$ = 1.7). Previous observations \citep{pet12} of the galaxies with the {\it Spitzer} InfraRed Spectrograph (IRS)  failed to detect spectroscopic evidence of an AGN from either of the Taffy galaxies. Formally, it has a hardness ratio (HR)\footnote{The hardness ratio (HR) is defined as H-S/H+S, where H and S are the background subtracted counts in the 0.5--2keV (S) and 2--8 keV(H) bands.} of~-0.06$\pm$0.11, the most positive of any source in the system (all the others have upper limits),  but lower than typical AGN (hardness ratios in the range 0.7--0.9).  

The elevated hardness ratio of W$_{nuc}$ may indicate that UGC~12914 contains a low luminosity AGN, accreting at a sub-Eddington rate \citep{ho09}.  Despite having only 90 net counts,  we have simulated how the HR varies over a range of power-law slopes and intrinsic hydrogen column densities.  Within the range allowed by a 2$\sigma$ uncertainty in HR, a typical source with  1.7~$<$~$\Gamma$ $<$~1.9, would require a range of possible absorbing columns of 0.3-1 $\times$ 10$^{22}$ cm$^{-2}$. This is consistent with an AGN with moderate intrinsic absorption. However, given the small number of detected counts at higher-energy, we cannot completely rule out a highly-obscured, Compton-thick AGN \citep[e.g.][]{kom03}. Although such a source in isolation would exhibit a much larger hardness ratio than observed, the HR might be lowered by the addition of a thermal component from extended nuclear star formation,  or shocks which would soften the spectrum as the AGN became more obscured. A conclusive test of whether UGC 12914 contains a Compton-thick AGN would require deep observations from a telescope sensitive to X-rays beyond 10~keV. 

It is reasonable to ask whether the compact sources are consistent with the X-ray Luminosity Function (XLF) of populations of compact X-ray sources found in other nearby systems.  For example, \citet{zez07} performed an in-depth study of $\sim$120 compact X-ray sources  discovered in the `Antennae' system \citep{fab03,zez06}, and was able to determine the XLF down to limits of 10$^{37}$ erg s$^{-1}$, a factor of $\sim$30 times fainter than the sources discovered in the Taffy. Their results were consistent with the brighter sources (ULXs) being an extension to high mass of a fainter population of X-ray binaries accreting close to their Eddington limits. 

Figure 4 shows the cumulative XLF determined for the compact sources (cyan line bracketed by black uncertainty boundaries). The luminosity function of the cumulative source number counts can be approximated by a power-law of the form  N($>$L) = AL$^{-\alpha}$ \citep[e.g.][]{kil02}. Here N($>$L) is the number of sources with luminosity greater than L, A is a normalization factor, and $\alpha$ is the slope of the assumed relationship. 
Following \citet{zez07}, we binned the luminosities of the 2 and 3-sigma sources in intervals of the natural unit of counts (one count converts closely to a unit of 1.25 $\times$ 10$^{38}$ ergs s$^{-1}$ at a distance to Taffy of 62 Mpc). We calculated the luminosities of the fainter compact sources in the same manner as the brighter sources (assuming a power-law spectral index of 1.7). The XLF shows a smooth transition from the brighter (3-sigma sources) to the fainter (2-sigma) sources, suggesting that they form a continuous population. Also plotted, to provide
additional information, are the differential counts as green and red squares in Figure 4. 

Given the small numbers of total sources (we only have 25 in all, and only 9 above 3$\sigma$--excluding W$_{nuc}$), we did not attempt to fit the slope of the XLF, nor did we attempt to correct for completeness in the fainter sources. Rather we simply compare our data with results from the deeper survey of Zezas et al. (2006) for the average $\alpha$ = 0.55 determined for the full deep `Antennae' survey (purple line).  This slope is dominated by the many fainter sources (L$_X$ $<$ 10$^{38}$ erg s$^{-1}$) in the `Antennae', which is characteristic of the massive X-ray binary population. 
It is clear from the comparison that Taffy sources appear to follow a steeper slope. However, although it might be tempting to speculate that the Taffy source may be powered by something other than massive X-ray binaries, the steepness is very likely a consequence of small number statistics in our sample. Indeed the same effect was found by Zezas et al. when they fit these same `Antennae' data with a split power law. These authors showed that sources brighter than L$_X$ $>$ 7.5 $\times$ 10$^{37}$ erg s$^{-1}$ (this would be the case for all our sources) could be fit with an $\alpha$ = 1.35 (red line). This slope is much more consistent with the Taffy sources. As Zezas et al. describe in their paper, the steeper slope is most likely an artifact poor statistical sampling at the bright end of the XLF, and the potentially rapid variability of fluxes found predominantly  in the higher-luminosity sources. Given, in our case, the additional uncertainty in the fluxes of the fainter sources, we conclude that our data are consistent with the sources being the tip of the luminosity function of high-mass X-ray binaries.


\subsection{X-ray Spectra and Modeling of Extended Regions} 

For the galaxy extractions and the bridge (the sum of extracted regions Bridge-1 and Bridge-2 of Figure 2a), we have enough counts to spectrally model these regions. Spectral modeling was done using the {\sc sherpa} packages of {\sc ciao}, having first removed the 3$\sigma$ point sources.  For the diffuse emission in the galaxies and bridge, we fit combinations of APEC thermal models \citep[Astrophysical Plasma Emission Code;][]{smi01} and a power-law. Solar metallicity is assumed.  APEC uses a model database of more than a million atomic spectral lines, plus free-free continuum emission for a hot plasma, to find the best fitting parameters consistent with these data.    A fixed Galactic foreground absorption $N_{H}=4.69\times10^{20}\,{\rm cm^{-2}}$ \citep{kal05}\footnote{\url{http://heasarc.nasa.gov/cgi-bin/Tools/w3nh/w3nh.pl}} is assumed. We also tried adding intrinsic absorption as a free parameter. The results are given in Table \ref{fits}. In addition to the name of the region (column 1) and the model type (column 2), Table \ref{fits} also provides the APEC normalization factor N$_{APEC}$ (column 3; proportional to the volume emissivity of the APEC--modeled plasma), thermal temperature kT (column 4), power-law flux density at 1 keV (column 5), power-law index $\Gamma$ (column 6), best-fitting intrinsic HI absorption column density--if allowed to be non-zero (column 7), and the normalized minimum $\chi^2$ of the final fit ($\chi^2$/n), where n is the number of degrees of freedom in the model.    

In Figure 5, we plot X-ray spectra from the observations, and over-plot the models.  
The spectra of the diffuse emission in UGC 12914 and UGC 12915  (Figure 5a and b) were best fit by a combined APEC and power-law fit.  For UGC 12914, we fitted a thermal component at 0.71($\pm$0.25)\,keV, and a power law model with a photon index of $\Gamma$ = 2.3($\pm$1.5).  Similarly, for UGC 12915, the spectrum is best fit with a  thermal model at 0.72($\pm$0.16)\,keV, and  a power law with a photon index of $\Gamma=2.0(\pm0.8$). We note that the differences in the normalized minimum $\chi^2$ for the fit, with and without a power-law, is not large. Nevertheless, we argue that inclusion of a power-law component in the modeling is physically reasonable, because an extrapolation of the XLF of the observed compact sources to fainter flux levels implies the existence of population of faint power-law sources which would appear as a diffuse component when averaged together\footnote{We have extrapolated the XLF of the compact sources down to L${_X}$ = 10$^{38}$ erg s$^{-1}$ and find that the expected unobserved luminosity from massive X-ray binaries is within a factor of two of the power-law component of the two galaxies measured from the spectral modeling.}. To improve the signal to noise of the modeling, we have also summed the signal from both galaxies, and re-fit the resulting spectrum (Figure 5c). As Table \ref{fits} shows, a significantly improved fit was found for the summed galaxies. We consider the APEC+Power-law fit to be the most reasonable with a new power-law index that is closer to that expected for the brighter point sources. Although formally the APEC+absorption provides an equally good fit, it is less compelling because it does not include the expected compact source contributions. The pure APEC model results in a fit of lower significance.  

In contrast, the spectrum of the emission from the entire bridge (Figure 5d) is best fit with a single thermal model of 0.63($\pm$0.17)\,keV with some intrinsic absorption. The amount of intrinsic absorption is plausible, given the gas-rich nature of the bridge.   Fits with and without absorption are shown in Figure 5d.   The inclusion of intrinsic absorption increases the 0.5--8\,keV X-ray luminosity of the bridge a factor of 2.6.  In conclusion, the temperature of the thermal component for the galaxies and the bridge are quite similar within the uncertainties, with values ranging between 0.7 and 0.8 keV  without absorption, and 0.5--0.7keV if intrinsic absorption is present.
 
Table \ref{model} present the derived properties of the hot gas. The mass of hot X-ray gas derived from the observations in the bridge is estimated to be $\sim$0.8--1.3 $\times$ 10$^8$ (f$_{fill}$)$^{0.5}$ M$_{\odot}$, where f$_{fill}$ is the filling factor of the X-rays.  Assumed a filling factor of unity for the X-ray emitting gas, this
is 11$\%$ of the total warm H$_2$ mass of 9$^{+2}_{-5}$ $\times$ 10$^{8}$ M$_{\odot}$,  extrapolated to the whole bridge by Peterson et al. (2012).     By comparison, the mass of dust estimated from the spectral energy distribution (SED) for region Bridge-1+ Bridge-2 is 7 $\times$ 10$^{7}$ M$_{\odot}$, or a total gas mass of 7 $\times$ 10$^9$ M$_{\odot}$ assuming a gas to dust ratio of 100. This is  close to to sum of the HI and H$_2$ masses in the bridge estimated by \citet{zhu07}.  The X-ray gas therefore constitutes $\sim$1$\%$ of the total gas mass in the bridge.

\section{Comparison of Extended Soft X-ray distribution to Dust and Gas distributions}

As shown in Figure 1b and c, the X-ray emission only correlates approximately with the radio continuum emission in the two galaxy disks, but less so in the bridge. The radio continuum emission in UGC 12915 correlates best with the collection of compact sources and the bright nucleus of UGC 12914. In the bridge, however, the X-ray emission appears to avoid the main radio continuum ridge and is distributed further to the north and west.  It is possible that the magnetic field is more strongly coupled to the denser molecular gas which also has a different distribution from the X-ray emitting gas (see below). The radio emission at $\lambda$6cm is strongly polarized and quite regular across the whole bridge  \citep{drz11} and shows no sign of Faraday depolarization in the direction of the X-rays, as commonly seen in radio galaxies immersed in hot X-ray gas \citep{gar91,har14}.    

In Figures 6a-6c, we show the extended X-ray emission compared with images of dust and gas from {\it Spitzer} and {\it Herschel}. The X-ray data have been color-coded according to energy, using adaptive smoothing after removal of the bright compact 3--$\sigma$ sources listed in Table 1 (see also Section 3.2).  Figure 6a shows contours of the 24$\mu$m emission from {\it Spitzer} of the galaxies superimposed on the X-ray color image.  In Figure 6b and c, we similarly over-plot  [CII]157.7$\mu$m line emission, and 100$\mu$m dust emission from the {\it Herschel} PACS spectroscopy and photometry data.  The [CII] line map was made by summing over the full line profile derived from a small map made with the PACS integral field unit spectrometer (Peterson \& Appleton in prep.). The extended soft X-rays seem more closely associated with the IR continuum emission than the brighter ridge of radio continuum. The [CII] emission, which represents the diffuse neutral medium unshielded from ambient UV light, is also quite well correlated with the X-rays.  

Figure 7a and 7b shows the soft X-ray emission superimposed on the integrated BIMA CO (1--0) molecular gas map and VLA HI map respectively. It is striking that the CO (1--0) molecular gas in the Taffy bridge seems anti-correlated with the soft X-rays, whereas the HI, like the dust and [CII] emission, is better correlated. In the models of \citet{vol12}, the offset between the diffuse HI and the denser CO--emitting gas was explained in terms of the impact parameter in the
collision. The X-ray emission seems to follow this trend, being more correlated with the diffuse components of the bridge than the denser colder molecular gas.  

Another interpretation of the anti-correlation of the soft X-ray emission with the CO (1--0) distribution may have to do with selective absorption of X-rays across the bridge by dense gas. In Table \ref{soft}, we show that Bridge-2 is not detected in 0.5--2 keV X-rays to a 3--$\sigma$ level.  Could heavy absorption in that part of the bridge extinguish the X-ray signature?   Based on the intensity of the CO (1--0) emission from \citet{gao03a} in that region, and using the unusually low value of X$_{CO}$ = 2.6 $\times$ 10$^{19}$ cm$^{-2}$ (K km s$^{-1}$)$^{-1}$ measured by \citet{zhu07} in the bridge, we estimate that the H$_2$ column density would be $>$ 1 $\times$ 10$^{21}$ cm$^{-2}$ in Bridge-2.  If we assumed that Bridge-2 is as bright as Bridge-1 in X-rays,  we have performed simulations  that allow us to estimate that we would need $\sim$2 $\times$ 10$^{21}$ cm$^{-2}$ of molecular hydrogen to extinguish the X-rays to the 3$\sigma$ level observed. It is therefore plausible that {\it if X-rays occupied the Bridge-2 region at a similar intensity to that observed in Bridge-1}, then we would not detect that emission in the presence of moderate intrinsic absorption.  If true, then the total X-ray emission from the entire bridge could be a factor of 2 larger than observed. Higher resolution CO observations would be needed to determine more precisely how clumpy the CO emission is, relative to the X-rays, and whether this explanation is viable, since our estimates of the H$_2$ column are based on low-resolution (9.8 arcsecs) BIMA observations of \citet{gao03a}. An important missing piece of information is the filling factor, between the dense molecular and the X-ray emitting gas. The above argument assumes a 100$\%$ filling factor, which is probably not realistic because it would imply far more molecular gas than is observed\footnote{A 100$\%$ filling factor with a column density of 2 $\times$ 10$^{21}$ cm $^{-2}$ implies a molecular mass in the  filament(1 x 0.3 arcmin$^2$) of 1.2 $\times$ 10$^{10}$ M$_{\odot}$. This is approximately five times the estimated H$_2$ mass in the bridge \citep[warm and cold H$_2$,][]{zhu07,pet12}, and a factor of two larger than the total gas mass, including HI. A lower filling factor would be required to bring these into approximate agreement, and this would significantly reduce the effect of the absorption of the X-rays.}.

\section{Origin of X-rays in the Galaxies and Bridge}

\subsection{X-rays from Star formation and Supernova Heating}

The existence of soft X-rays in galaxies is a natural consequence of supernovae and winds from regions of star formation which create the hot thermal component which pervades the ISM \citep{cox79,cox81,mck77}. With advances in X-ray imaging, it was soon discovered that this hot gas can escape into the haloes of galaxies, especially in starburst systems where ionized gas and X-rays were seen along the minor axis of such galaxies \citep{wat84,fab84,fab88,hec90,fab90,arm95,rea97,dah98}. 
The vertical extent, surface brightness, and other properties of X-ray gas in edge-on normal and starburst galaxies was studied in detail by  \citet{str02,str04a,str04b}. These studies were able to explore the relationship between disk and halo X-ray gas and properties, and found that, in many cases, the vertical distribution of X-ray gas in these disk systems was an exponential with scale-heights of  typically 2--4 kpc. In low star-forming systems (with SFRs $<$ 0.5 M$_{\odot}$ yr$^{-1}$), gas was not found in the haloes, whereas in galaxies with significantly  higher star formation activity (3 $<$ SFR (M$_{\odot}$ yr$^{-1}$) $<$ 10), all had X-ray haloes. The surface brightness of the halos was typically a factor of 10 lower than in the disks, as was the local ratio of X-ray to far-IR luminosity in the haloes. 

With the advent of multi-wavelength archival data from the UV to the far IR, we are now in a position to accurately measure the star formation rates of the Taffy galaxies and bridge.  We  can then ask how normal is the observed X-ray emission in the galaxies, and specifically whether star formation alone could explain the existence of soft X-ray emission in the bridge?  

To measure the star formation rates for the galaxies and the bridge region, we use an SED modeling code MAGPHYS \citep{dac08,dac10}. This code fits the observed SEDs with sets of model templates, and estimates the likelihood distributions of physical parameters of galaxies, including SFRs, stellar mass, and IR luminosities.  
The SEDs were constructed for the two galaxies and bridge region (defined in Figure 2), utilizing all the available multi-wavelength imaging from UV ({\it GALEX}) to sub-mm (JCMT) data available
from the archives.  Figure 8 presents these SEDs for the two galaxies and for Bridge-1 (of Figure 2), along with fits using MAGPHYS. The derived properties of the two galaxies and the bridge regions, which are well fit by the models, are presented in Table \ref{ir}\footnote{The star formation rates may be overestimated because an unknown fraction of the far-IR emission may
come from re-processed scattered UV light from an older stellar population (e. g. Cirrus emission), which is not fully taken into account in MAGPHYS.}. 

The Taffy galaxies were previously known to be relatively weak IR emitters \citep{con93, jar99} compared with other interacting systems. The pair was detected (but barely resolved) by IRAS, and \citet{san03} quote a far-IR luminosity of 6.5 $\times$ 10$^{10}$ L$_{\odot}$.  Our results, using the full SED, are consistent with this. We find that UGC 12914 and UGC 12915 have far-IR luminosities of 1.0 and 3.5 $\times$ 10$^{10}$ L$_{\odot}$, and global SFRs of  1.05$^{+0.08}_{-0.13}$  and 2.6$^{+0.24}_{-1.10}$ M$_{\odot}$ yr$^{-1}$ respectively, where the uncertainty is derived from the range of allowable values found by fitting different templates in MAGPHYS.  \citet{jar99} estimate the star formation rate for the entire system to be $\sim$2--4 M$_{\odot}$ yr$^{-1}$, which is very similar to our values.\footnote{We do not confirm the unusually high star formation rates for the Taffy galaxies (7 and 14 M$_{\odot}$ yr$^{-1}$ for UCG 12914/5 respectively) obtained by \citet{kom12} using Pa--$\alpha$ emission-line fluxes.  Their measurements assume a very high optical extinction, and it is possible that the measured Pa--$\alpha$ fluxes are biased upwards by shock excitation of the gas in the disks \citep[see][]{pet12}. Indeed we find that the values quoted by \citet{kom12} do not follow the usual Pa--$\alpha$ to 24$\mu$m relationship observed by \citet{cal07}, perhaps supporting the idea that the Pa--$\alpha$ is not all coming from star formation. } 

It is known that the X-ray luminosity of a galaxy is correlated with the star formation rate \citep{fab85,fab89}.  Measuring the star formation rates of the Taffy galaxies using the soft X-ray luminosity calibrated for normal galaxies by \citet{min12b} of L$_{X(0.5--2keV)}$ erg s$^{-1}$  =   
8.3 $\times$ 10$^{38}$ $\times$ SFR M$_{\odot}$ yr$^{-1}$, yields much higher SFRs than those determined from the SED/IR methods. Based on the APEC luminosities in the 0.5-2keV range, we estimate SFR$_{X}$ = 10$^{+12}_{-5.4}$ and 4.7$^{+5.6}_{-2.5}$ M$_{\odot}$ yr$^{-1}$ respectively for UGC 12914 and UGC 12915. These rates (even taking into account the large uncertainty of 0.34 dex in the relationship) are not only higher than the SED--derived SFRs, but if taken at face value, imply the SFR is larger in UGC 12914 than UGC 12915. This is contrary to all the other previous published work. 
We believe that the most likely explanation for the discrepancy is that unlike normal galaxies, the Taffy galaxies contain significant quantities of gas shock-heated in the collision, rather than all originating from supernova heating. We already noted the odd distribution of hot gas in UGC 12914, suggesting that much of the emission may be from bridge-gas, and this may explain why this system is more luminous in soft X-rays than its companion.   This picture is supported by the compact sources (Section 3.2), where the number of very luminous X-ray point sources is larger in UGC 12915 than UGC12914, suggesting that the former, not the latter galaxy, is the more prolific of the pair\footnote{If we extend the analysis to fainter compact sources using N$_{XRB}$($>$ 10$^{38}$ erg s$^{-1}$) $\sim$ 3.22 SFR (M$_{\odot}$ yr$^{-1}$) from \citet[][]{min12a}, we find SFRs of 3.72$\pm$1.7 and 2.48$\pm$1.1 M$_{\odot}$ yr$^{-1}$ for UGC~12914 and UGC~12915 respectively. However, within the considerable uncertainties in this relationship, the higher SFR for UGC~12914 over than of UGC~12915 derived from fainter sources is not significant.}.

Adopting our SED--derived star formation rates, we compare the results with other systems that have supernovae-driven outflows. Only UCG 12915 has a global star formation rate comparable with galaxies like NGC 4631 and NGC 3628, which have known X-ray halos \citep{str04a}. In Figure 9, we show two cross-sections through the Taffy system approximately perpendicular to the major axes of both galaxies. Section A in Figure 9, passes through the nuclear regions of both galaxies, whereas Section B is offset from the centers but passes through the brightest parts of the bridge. These cross-sections show that the soft X-ray emission has a complex distribution, and does not follow a simple exponential profile expected if both galaxies had winds. UGC 12915, the galaxy one might expect to have a starburst wind, has a distribution that is strongly biased towards the bridge region in both cross-sections.  In Section A, UGC 12914  has an X-ray distribution that falls symmetrically about its nucleus, until it reaches the bridge where it remains strong. In Section B, clumps of soft emission are seen in either side of the disk of UGC 12914, but the emission again runs into a distinct bridge structure. Thus the bridge X-ray emission is clearly not a simple symmetrical starburst-driven halo.  

Returning to the bridge, we also find the average star formation rate across  region Bridge-1\footnote{As a check on the method, we also compare the SFR for Bridge-1 by estimating it from the 24$\mu$m emission alone and using the method of Calzetti et al. (2007). This provided a SFR of 0.24 M$_{\odot}$ yr$^{-1}$, which is very consistent with the full SED approach.} to be 0.26$^{+0.02}_{-0.04}$ M$_{\odot}$ yr$^{-1}$, and a slightly lower value of 0.19$^{+0.02}_{-0.01}$ M$_{\odot}$ yr$^{-1}$ for Bridge-2. These values are more than 10 times lower than the star formation rates in the disk of UGC 12915. Given that in normal galaxies, the X-ray luminosity is expected to scale roughly as SFR$^{1.08}$ \citep{ran03}, then
if the bridge X-rays were created in a normal galactic environment, they should be at least ten times fainter.  However, as Table \ref{soft} shows, the soft X-ray surface brightness in the Bridge-1 region is only a factor of two to three times less than the disk surface brightness. Alternatively, we can predict the expected star formation rate in the bridge based on the correlation derived for normal galaxies by \citet{min12b}. Based on this prediction, if star formation dominated the bridge X-ray emission, it  should have a SFR of  $\sim$ 2.8 M$_{\odot}$ yr$^{-1}$, whereas we measure it to be ten times smaller.  Therefore normal {\it in situ} star formation in the bridge is unlikely to be the origin of the X-rays. 

\subsection{Shocks and Turbulence in the Taffy Bridge}

Observations with {\it Spitzer} \citep{pet12}  revealed the existence of an  unusually large amount of warm molecular hydrogen
in the bridge. This was attributed to mechanical heating of the molecular gas by turbulent energy dissipation left over from the direct collision of the two galaxies. 
However, before discussing shocks further, it is worth ruling out a more mundane explanation for the abundance of
warm H$_2$:  that the molecular gas is heated because it is bathed in a strong X-ray field.  \citet{pet12} could not rule this out, since until the present work, no previous X-ray observations had been made. 

We estimate the X-ray luminosity in the bridge evaluated over the same areas as those observed by the Long-Low slit of the {\it Spitzer} IRS.  These measurements are then compared with the H$_2$ line luminosity L(H$_2$)$_{(0,1)}$, where the (0,1) designation refers to the 0-0 S(0)28$\mu$m and 0-0 S(1)17$\mu$m pure rotational lines of warm molecular hydrogen. This luminosity is almost certainly a lower limit to the total power emitted in the rotational hydrogen lines, since higher order transitions where not observed (they fell outside the wavelength range of the Long-Low grating). Summing the line luminosities over bridge regions B,C,D,E,J,K,L,M \citep[as defined by][]{pet12},  we find L(H$_2$)$_{(0,1)}$= 1.2 $\times$ 10$^{41}$ erg s$^{-1}$.  The X-ray luminosity over the same area is found to be L$_{X(0.5-2keV)}$ = 2.7 $\times$10$^{39}$ erg s$^{-1}$. This implies that  L(H$_2$)$_{warm}$/L$_X$ $>$ 42. Given the inherently inefficient ($<$ 1$\%$) heating of H$_2$ by X-rays, this high ratio rules out X-ray heating of cold molecular gas as a mechanism to explain the high luminosity of warm H$_2$.  The interpretation of the {\it Spitzer} observations in terms of turbulence and shocks by \citet{pet12} is therefore upheld. 

If the X-ray emitting gas is heated by shocks, what does the temperature of the bridge gas imply about the shock velocity? Based on the spectral fitting of the bridge, our best model for the Bridge-1 region yields a temperature  kT = 0.63 keV (T=7.3 $\times$ 10$^6$ K) with absorption, or 0.77 keV (8.9 $\times$ 10$^6$ K) in the absence of significant absorption. For an idealized radiationless J-shock driven perpendicularly into a low-density perfect gas with high Mach number, the temperature of the post-shock gas, T$_{ps}$= (3$\mu$m$_H$/16k) $v_s$$^2$, where $\mu$ is the mean molecular weight, m$_H$ is the mass of the hydrogen atom, k is the Bolztmann constant, and $v_s$ is the shock velocity. For the range of temperatures we obtained from the {\it Chandra} observations, shocks with velocities of  430 $<$ $v_s$ $<$ 570 km s$^{-1}$ would be required if the pre-shocked medium was neutral. Such velocities are not out of the question, as it has been estimated that the Taffy galaxies collided with an impact velocity of $\sim$ 600 km s$^{-1}$ \citep{con93}. When combining this with the high rotational--velocity retrograde collision \citep{bra03}, the impact velocities at the shock interface could have been as high as 800 km s$^{-1}$.  The lower-mass UGC 12915 could collide at velocities $>$500 km s$^{-1}$ in the center of mass frame. Despite the uncertainties in the actual relative velocities of the two galaxies (most of the motion is now in the plane of the sky), the heating of the lower-density ISM in these galaxies to high temperatures is quite plausible.  

Hot gas heated in the original impact exceeding 5 $\times$ 10$^{6}$ K (0.43 keV), will cool extremely slowly \citep[few $\times$~10$^7$ to 10$^8$ years;][]{gui09}\footnote{The cooling times for the hot gas in the Bridge and the Galaxies based on simple energy arguments (Table \ref{model}) are of order 1Gyr, but unlike Guillard et al.~(2009), do not include the important cooling effects of the spluttering of small grains which considerably reduces the X-ray gas cooling time. Effects of heat conduction in supersonic turbulence (not included) would further reduce the cooling time. }: longer than the time since collision of the two disks, based on radio synchrotron aging in the bridge, is 2~$\times$ 10$^7$ yr \citep{con93}.  It is therefore likely that the X-ray emitting bridge gas is a cooling remnant of the pre-shocked low-density material,  splashed out into the bridge that has not yet cooled down.

Evidence that gas has been splashed out may be seen not just in the bridge, but also in the distribution of hot gas in the galaxies.  Although in UGC 12914, the soft X-rays are well correlated with the nucleus and the northern half of the galaxy, they show a significant absence of X-rays in the southern half of the disk. For a normal galaxy, this would be unexpected, since that region of the galaxy is populated with both warm and cool dust, and diffuse gas (Figure 6). A natural explanation is  the X-ray emission from UGC 12914 is entirely dominated by shocked gas in the bridge (seen in projection against the northern disk), and the intrinsic soft X-ray emission from UGC 12914 is too faint to be detected. Although UGC 12915 shows a closer correlation between the X-ray emission and the brighter
parts of the molecular disk, there are strong deviations along the north-western part of the disk where the CO emission is weak but X-rays are strong. This may again indicate that the disk gas is strong disturbed in the direction of the bridge emission. 

The reality of the physical conditions in the Taffy bridge are likely to be much more complex than that of an idealized perfect gas experiencing shocks. Numerical hydrodynamic simulations of a head-on collision of this kind by Struck (1993), predicted that shortly after the collision, a "splash bridge" should form containing a mix of hot gas from the original collision, with cooler gas reforming rapidly.  \citet{vol12} showed that with the right distribution of molecular cloud sizes and collisional impact parameters, significant amounts of both dense and diffuse gas can be stripped from the galaxies into the bridge. Their model was able to  reproduce many of the observed morphological and kinematic features of the bridge.  It also provided an explanation for the offset spatial distributions of dense CO-emitting gas, and the diffuse HI. In the model, this resulted from UGC 12914 striking the disk of UGC 12915 significantly
off--center, leading to an enhancement of diffuse HI in the western part of the bridge compared with the denser eastern part. 

This is consistent with our suggestion that the X-ray emitting gas 
is tracing the path of the lower-density material in the pre-collisional disk of UGC 12914, drawn-out and offset from the denser gas as a result of the collision geometry. Although the \citet{vol12} models seem to reproduce the basic distribution and kinematics of the gas very well, this "sticky-particle" modeling approach, like most current hydrodynamic models, is unable to properly deal with the obvious multi-phase nature of the gas in the bridge. The fact that gas can be both ejected from the galaxies, and can also rapidly cool in the aftermath of the collision, means that a proper description of the relative masses in each phase must take into account transformations (and mixing) between the phases, and detailed gas cooling in shocks and turbulence. 

By analogy with the Taffy bridge, \citet{gui09} presented a multi-phase model to explain X-ray (Trinchieri et al. 2003; 2005, O'Sullivan 2009) and warm H$_2$ emission (Appleton et al. 2006, Cluver et al. 2010) in the intergalactic shock in Stephan's Quintet. This model was designed to treat the case of a high-speed (1000 km s$^{-1}$)  collision of an intruder galaxy, NGC~7318b with an large 40 kpc-long HI tidal filament. The model showed that the initial high-speed shock moving through multi-phase HI filament would shock--heat the low-density pockets to X-ray temperatures, rapidly spluttering the dust in those regions, while the denser pre-shock  gas could survive because it is shocked to lower velocities. 

Like Stephan's Quintet, the pressure (P/k) of the hot  diffuse X-ray emitting gas in the Taffy bridge (see Table \ref{model}) $\sim$10$^4$ K cm$^{-3}$, is fifty times less than the pressure in the warm, dense H$_2$ phase of bridge \citep[$\sim$6$\times$10$^5$ K cm$^{-3}$ for T= 165 K, and n(H$_2$)=10$^3$;][]{pet12}. As discussed by Guillard et al. (2009), supersonic turbulence leads to significant pressure differences in the post-shock gas, with no expectation of pressure equilibrium. Indeed, the cooling time for the warm H$_2$ is so short that warm H$_2$ must be continuously created, as individual dense regions dissolve as they expand into the ambient hot gas. This implies rapid cycling of material through different ISM phases. 

The ratio of the warm H$_2$ to X-ray luminosity L(H$_2$)/L$_X$ in Stephan's Quintet ($\sim$5--10) is much smaller than that seen in the Taffy bridge, L(H$_2$)/L$_X$ $>$ 42. 
The L(H$_2$)/L$_X$ ratio for Taffy is also particularly large compared with other shocked systems \citep[see Figure 9 of][]{lan15}. Part of the explanation may relate to the difference in thermal energy imparted by the faster collision in the Quintet (the difference in imparted kinetic energy could be a factor of 5) compared with Taffy.  The difference could also relate the distribution of pre-shock densities. In the Quintet's shock,  it is believed that the
intruder collided with a diffuse HI filament, not a complete galaxy disk, thus increasing the amount of low-density material available for X-ray heating. In the Taffy, there may have been a much larger fraction of dense gas present in both colliders, which would enhance the warm H$_2$ fraction relative to the hot X-rays. 


\section{Why is the Taffy system not a strong IR-emittor?}

Previous observations of the Taffy system have shown that the bridge contains $\sim$7$\times$ 10$^9$ M$_{\odot}$ of molecular and atomic gas \citep{zhu07}.  The existence of the X-ray emission adds to the multi-phase nature of the bridge, and supports the view that  the bridge material was gas stripped away from the galaxies in the head-on collision. The stripping of gas from the galaxies, and the turbulent nature of the bridge gas, may partly explain why the galaxies have not yet evolved into LIRG (the combined IR luminosity of the Taffy galaxies, L$_{FIR}$ = 4.5 $\times$ 10$^{10}$ L$_{\odot}$; see Table \ref{ir}).

For the bridge, the depletion timescale, $t_{dep}$, defined as the total gas mass divided by the star formation rate,  is $\sim$30 Gyrs, assuming an average SFR of 0.225 M$_{\odot}$ yr$^{-1}$ (see Table \ref{ir}). If only molecular gas (cold and warm) is considered in the estimate \citep[M(H$_2$) = 2.3 $\times$ 10$^{9}$ M$_{\odot}$ for D = 62 Mpc,][]{zhu07,pet12}, then $t_{del}$ $\sim$10 Gyrs.   These values\footnote{The adopted value for $X_{CO}$ (the ratio of N$_{H2}$ to $I_{CO(1-0)}$) for the bridge, used by Zhu et al. (2007) of 2.6 $\times$ 10$^{-19}$ cm$^{-2}$ K km s$^{-1}$, is very conservative, being a factor of ten lower than that adopted for the Galaxy, based on their LVG excitation analysis. Thus the depletion times could be underestimated if the total gas mass is larger than assumed here.}, which are comparable with, or greater than the age of the universe,  
suggest very inefficient star formation in the bridge (the star formation efficiency is the inverse of $t_{del}$).  A similar conclusion, that the bridge exhibits unusually low star formation for its gas mass, was found by \citet{vol12} using {\it Spitzer} data alone. Long depletion times have been found in a sample of warm H$_2$--rich Hickson Compact Group galaxies (Alatalo et al. submitted), and in NGC 1266 \citep{ala15}. In both situations, shock-heating of the gas was implicated in the star formation suppression. In Taffy, the depletion times are likely to be lower limits, because the star formation rate in the bridge is almost certainly overestimated, since the SED fitting assumes the UV radiation from young stars is the main heating mechanism for the dust. 

We can also estimate the gas depletion times for the individual galaxies. Adopting the total H$_2$ +HI masses for UGC 12914 and UGC 12915 of 9.2 and 8.2 $\times$ 10$^9$ M$_{\odot}$ \citep{zhu07}, and the star formation rates from Table \ref{ir}, we find t$_{del}$ = 8.8 and 3.2 Gyrs respectively. 
If we only include the H$_2$ masses, the depletion times drop to 4 and 1.6 Gyrs respectively (or log$_{10}$($t_{del}$) = 9.6 and 9.2).  \citet{hua14} have shown that, for normal galaxies, there is a good correlation between
H$_2$--derived log($t_{del}$) and stellar mass log(M$_{*}$). If we adopt the values of stellar mass from the SED fitting, the correlation of \citet{hua14} would predict depletion times of log(t$_{del}$) = 9.1 and 9.0 respectively for UGC 12914 and UGC 12915 with a 1-$\sigma$ scatter of dex 0.27.   The observed value of log($t_{del}$) for UGC 12915 is therefore well within the main distribution for normal galaxies, and UGC12914 deviates by only 1.9$\sigma$ from the mean relation (on the star formation deficient side). We conclude that, unlike the bridge, the Taffy galaxies have depletion times (and star formation efficiencies) that are within the normal range for their gas content. Nevertheless, their $t_{del}$ values are an order of magnitude longer than those of LIRGs \citep{zhu07}.  

The most likely reason that the Taffy galaxies have not yet become strong IR emitters, despite having recently undergone a major gas-rich collision, is the geometry of the head-on collision. This  has strongly disrupted the gas, splashing it \citep{str97} into the bridge where much of it is still turbulent. The bridge gas still contains large quantities of warm H$_2$, \citep[see][]{pet12}, as well as a significant component of hot X-ray emitting gas. In addition, there has not been enough time since the collision (25 Myrs) for the bridge gas to fall back onto the disks, which might eventually lead to more activity. 

\citet{jar99} pointed out the existence of a ring structure in UGC 12914, probably arising from the same process that leads to collisional ring galaxies \citep[see][]{app96}. The geometry of the Taffy system is such that both galaxies might be expected to form radially expanding rings within their disks in this nearly face-on collision.  It is possible that the ULXs in UGC 12915 are part of a forming gas-rich ring, seen edge-on. For example, in \citet{gao03b}, more than a dozen ULXs were discovered associated with the main expanding star forming ring of the Cartwheel collisional galaxy. As \citet{app87} showed, 
collisional ring galaxies are rarely bright enough to be LIRGs, partly because the star formation is confined to a narrow ring (or multiple rings), and generally does not form a strong nuclear starburst.  We conclude that the peculiar near-central collisional geometry, combined with strong gas stripping, has predisposed the Taffy galaxies to be less active than a more typical major merger.

\section{Conclusions}

The X-ray observations of the Taffy galaxies and bridge, combined with new {\it Herschel} data,  have shown the following main results: 

\begin{itemize}

\item X-ray emission is detected from both galaxies and from the bridge regions of the Taffy system. The galaxies themselves contain both compact point-like sources and extended soft X-ray emission. 

\item Nine ULX (L$_X$ $\sim$ 10$^{39}$ erg s$^{-1}$), and potentially more fainter point-like sources, are found in the Taffy system, and the brightest ULX (L$_{X(0.5-8keV)}$ = 2.8 $\times$10$^{39}$ erg s$^{-1}$) lies in the bridge, close to the position of a giant extragalactic HII region which is associated with a peak in the molecular gas density in the bridge. An additional, slightly extended source with a harder X-ray spectrum, is found associated with the nucleus of UGC 12914, perhaps hinting at the existence of a buried AGN. The number of ULXs in the two galaxies scales roughly with the SFR, as found in previous studies, suggesting that these X-ray sources are generated in massive X-ray binary systems, with an evolution that is intimately tied to massive star formation.   This is consistent with the shape of the XLF for the compact sources. 

\item Extended X-ray emission, L$_{X(0.5-8keV)}$  =  5.4 $\times$ 10$^{39}$ erg s$^{-1}$, is found primarily in the north-western part of the bridge. The total mass in hot gas is (0.8--1.3)~$\times$~10$^8$~M$_{\odot}$, which is approximately 1$\%$ of the total (HI~+~H$_2$) gas mass in the bridge, and $\sim$11$\%$ of the mass of warm molecular hydrogen discovered by{\it~Spitzer}. It is more closely associated with far-IR emitting diffuse gas and dust, than with either the brightest ridge of radio continuum emission seen in the bridge or with the distribution of denser gas tracers in the bridge. The lack of obvious correlation with the brighter radio continuum emission in the bridge rules out a direct connection between the X-rays and the synchrotron emission in the bridge caused by cosmic rays.   

\item  We show, using full SED fitting including {\it Herschel} far-IR data, that the star formation rate (0.24 M$_{\odot}$ yr$^{-1}$), is too low in the bridge to explain the relatively high X-ray surface brightness detected there. Furthermore, only UGC 12915 has a global star formation rate comparable with previously studied minor-axis X-ray outflow galaxies, and the X-ray surface brightness profiles are quite dissimilar to galaxies with known X-ray outflows. We favor shock heating from the galaxy collision as the main source of X-rays. The offset distribution of the diffuse gas tracers (including the X-ray emission) from the dense gas tracers  in the bridge, may be the results of an off-center collision between the two galaxies. 

\item  
We suggest that the shock-heating of low-density material in the precursor ISM of both galaxies has been splashed out \citep[see][]{str97}
into the bridge along with other diffuse gas leaving a highly turbulent bridge unable to support much star formation. The properties of the X-ray bridge implies that the gas is only just beginning to cool significantly in the 25 Myrs since the collision when it was likely first shocked. The temperature of the extended soft X-ray emitting gas is consistent with heating by shocks of over a range of 430 to 570 km$^{-1}$ depending on the assumed intrinsic absorption in the X-ray spectral modeling. These values are similar to the impact velocities assumed for the head-on collision that has led to the bridge formation.  The pressure in the hot gas is found to be much lower than the implied pressure in the warm H$_2$, as expected from models of supersonic turbulence which predict rapid cycling of molecular material through many different phases in the post-shocked region. 

\item The luminosity of the X-ray emission is 42 times lower than the mid-IR line--luminosity of warm H$_2$ emission seen in the bridge by {\it Spitzer} in previous observations. The new {\it Chandra} observations thus rule out the X-rays themselves as being the heat-source for the 10$^9$ M$_{\odot}$  of warm molecular gas in energetic grounds. Exploring the importance of  X-ray heating of the molecular gas in the bridge was an important motivation for this study. 

\item The combination of {\it Herschel} photometry, with full SED fitting, implies normal gas depletion times for the two Taffy galaxies (implying normal star formation efficiencies) but an extremely long depletion time ($>$10 Gyrs) for the gas bridge. Star formation therefore seems suppressed in the bridge where warm H$_2$ and X-ray emitting gas have not yet cooled, and velocity dispersions remain high in the molecular gas. We suggest that
the reason the Taffy system is not a strong IR-emittor (as with many other major gas-rich mergers) is related to the collisional geometry, which has stripped and disrupted a large amount of the gas into the bridge.   

\end{itemize}

\acknowledgements
We thank an anonymous referee for very helpful comments that improved the manuscript. The scientific results reported in this article are based on observations made by the Chandra X-ray Observatory.
Support for this work was provided by the National Aeronautics and Space Administration through Chandra Award Number GO3-14087X, issued by the Chandra X-ray Observatory Center, which is operated by the Smithsonian Astrophysical Observatory for and on behalf of the National Aeronautics Space Administration under contract NAS8-03060. The work is also based, in part, on observations made with {\it Herschel}, a European Space Agency Cornerstone Mission with significant participation by NASA. Partial support for LL, PA \& KA was also provided for {\it Herschel} observations, through a contract issued by the Jet Propulsion Laboratory, California Institute of Technology under a contract with NASA. This research has made use of the NASA/ IPAC Infrared Science Archive, which is operated by the Jet Propulsion Laboratory, California Institute of Technology, under contract with the National Aeronautics and Space Administration. Partial support for KA was also provided by NASA through Hubble Fellowship grant \hbox{\#HST-HF2-51352.001} awarded by the Space Telescope Science Institute, which is operated by the Association of Universities for Research in Astronomy, Inc., for NASA, under contract NAS5-26555. J.W. acknowledges support from NSFC grants 11443003 and 11473021.

\newpage

\begin{figure*}
\centerline{\includegraphics[width=7.5 in]{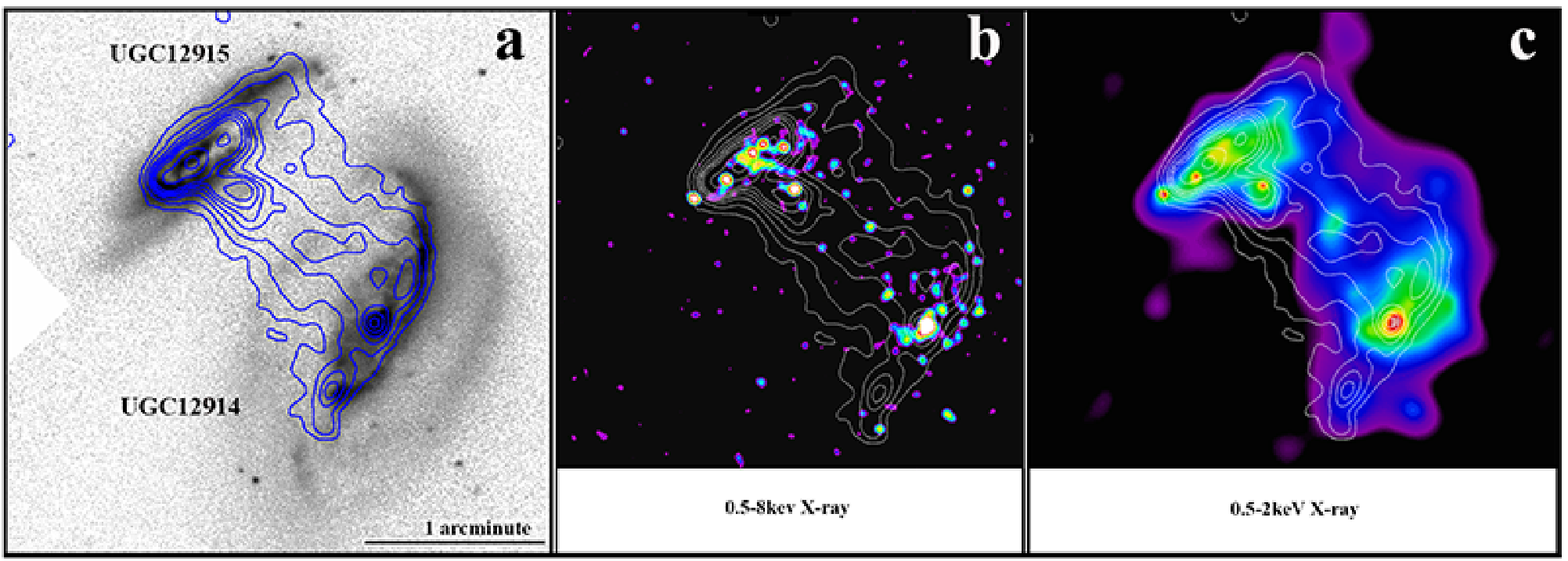}}
\caption{ (a) {\it VLA} 20 cm radio continuum map of Taffy system superimposed on g-band SDSS image. Contour units are 0.1, 0.3, 0.6, 0.9, 1.2, 1.5, 2, 4, 7 mJy beam$^{-1}$ \citep[adapted from][] {con93}, (b) 0.5--8 keV {\it Chandra} X-ray
image slighty smoothed to emphasize the compact sources, and (c) 0.5--2 keV (soft) X-ray band adaptively smoothed (see text) to emphasize the faint extended soft X-rays from the bridge (contribution from point sources are included--but see Figure 2). Note the offset between the peak radio continuum contours in the bridge (same units as in Fig.1a) and the brightest ridge in the extended soft X-ray emission. }
\end{figure*}

\begin{figure*}
\centerline{\includegraphics[width=7.5in]{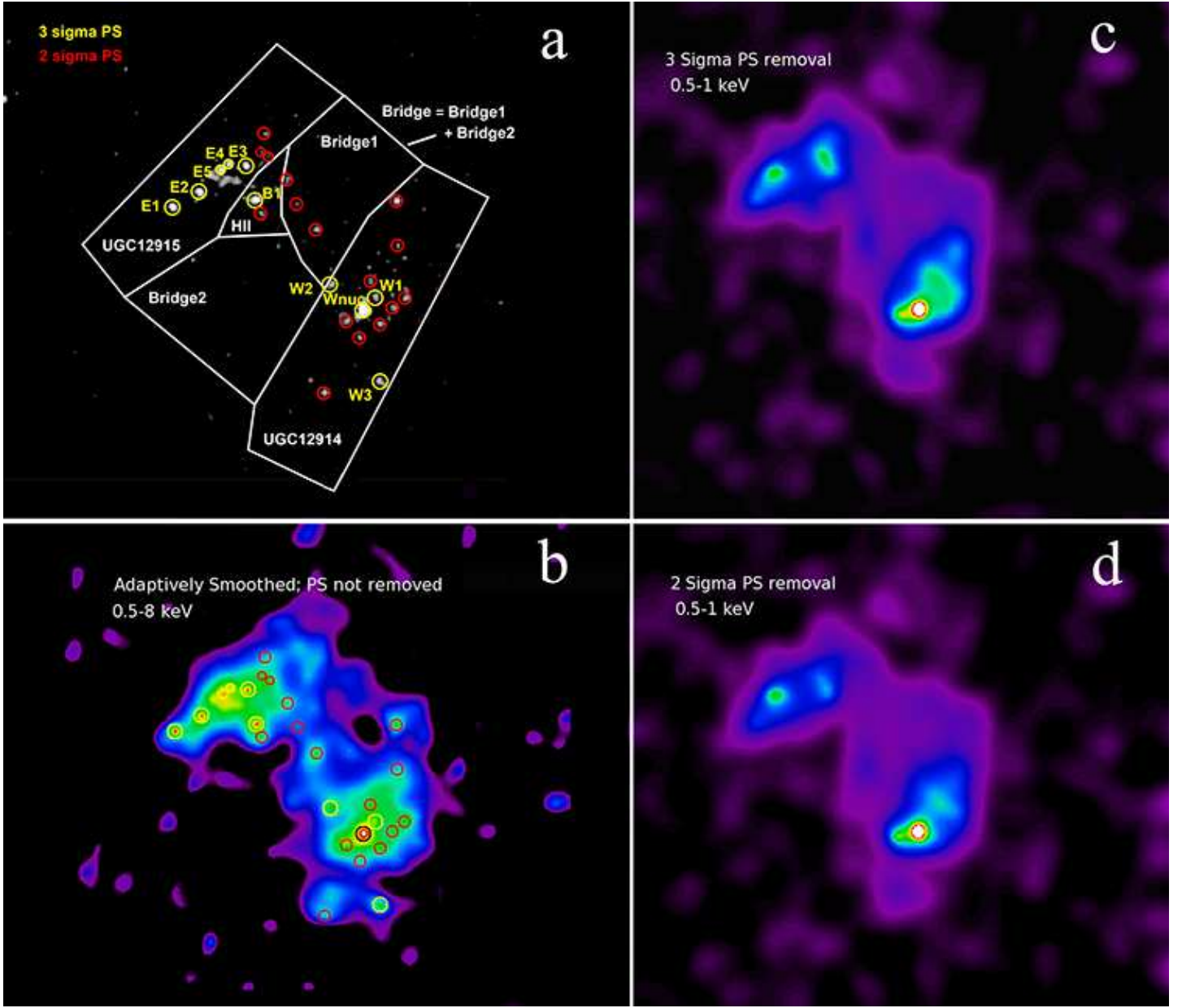}}
\caption{ (a) Extraction regions for both bright point-like sources W1-3, E1-5, and B1 associated with the south-western galaxy UGC 12914, north-eastern galaxy UGC 12915, and the bridge region respectively; and the extended regions
extracted (see Table 1). The bridge region is split into two parts to show the south-east, north-west asymmetry in surface brightness discussed in the text, and excludes the region associated with the bright HII region. The HII region (triangular-shaped extraction region) contains the strong ULX B1, but is especially dominant at other wavebands. (b) Point sources overlaid on the adaptively smoothed 0.5--8 keV map.  W$_{nuc}$ appears to be the nucleus of UGC 12914 (when compared with archival 2MASS K-band imaging and radio emission), whereas UGC 12915 does not have an obvious single nuclear compact X-ray source,  (c) Soft X-ray adaptively smoothed images with the removal of the 3$\sigma$ as well as (d) 2$\sigma$ sources (see text). We have not removed the possible nuclear sources E5 and W$_{nuc}$ from these images.  }
\end{figure*}

\newpage

\begin{figure*}

\centerline{\includegraphics[width=5in]{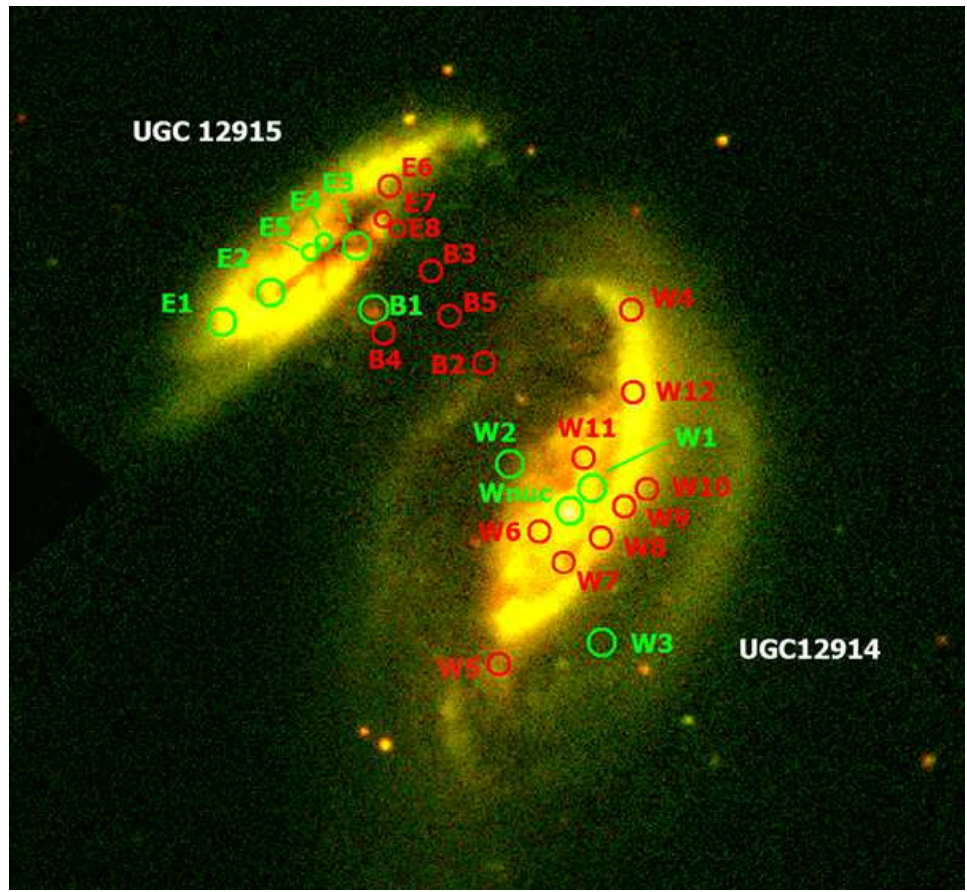}}
\caption{Distribution of compact sources in the Taffy system superimposed on the SDSS {\it ugi} color image. The red circles show the positions of the fainter 2-sigma sources (Table \ref{compact}), and the green sources show the 3-sigma sources (Table 1). Based on their count rates within the detection apertures,  we expect no more than one faint source per Taffy field
to be a chance association with a background source. Note that most of the compact sources are loosely associated with structure in the galaxies. Sources B1-B5 appear to be distributed within the bridge (and possible faint dust bridge).  B1 is associated with the known extragalactic HII region
and star cluster and probably formed recently within the bridge gas, as it is associated with several bright clumps of molecular hydrogen. }
\end{figure*}

\newpage

\begin{figure*}
\centerline{\includegraphics[width=6in]{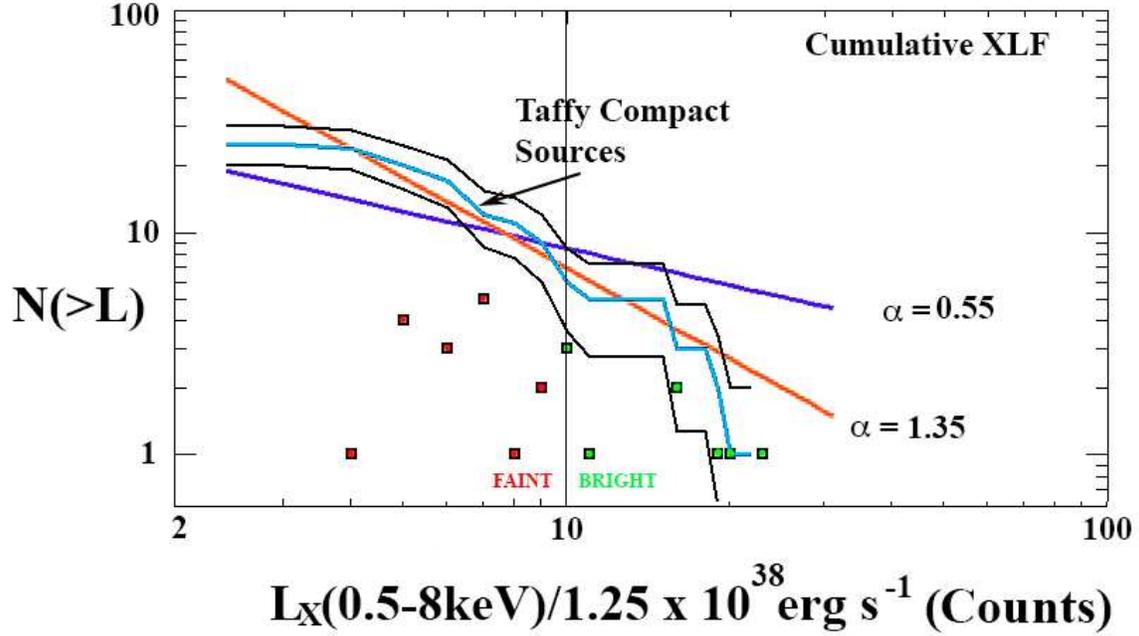}}
\caption{The cumulative X-ray Luminosity Function (XLF) of the compact sources in the Taffy system (cyan line) with uncertainties (black lines) based on Poisson noise produced by the binning only. The purple line
shows the slope from Zezas et al. (2006) for the complete `Antennae' sample of compact sources, and the red line shows the slope derived by Zezas et al.  for the bright end of the same `Antennae' sample, when bright and faint sources are fit with different slopes. For completeness we also plot the differential number counts in the same bins (red squares represent the fainter 2-sigma sources, and the green squares show the counts for the brighter 3-sigma sources; see text).}
\end{figure*}
\newpage

\newpage

\begin{figure*}
\centerline{\includegraphics[height=8in]{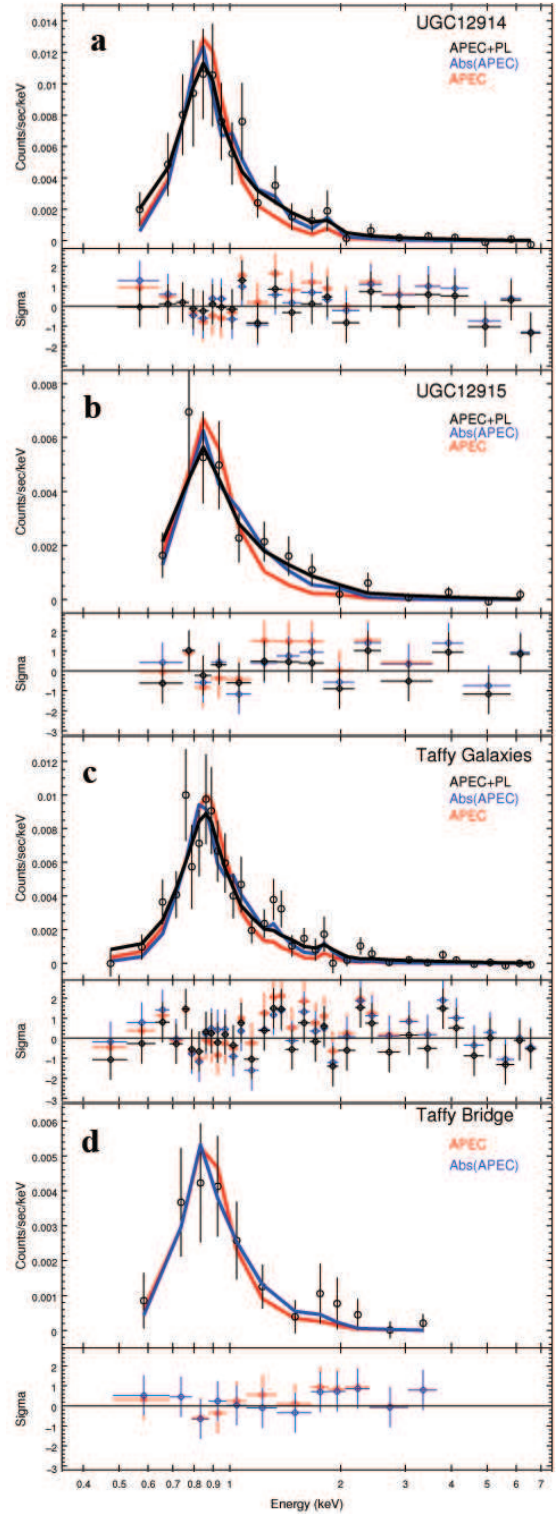}}
\caption{X-ray spectra of (a) UGC 12914, (b) UGC12915, (c) both galaxies summed together, and (d) the bridge, with best-fitting models, and the residuals in the lower panel of each subfigure.}
\end{figure*}

\pagebreak

\begin{figure*}
\centerline{\includegraphics[width=7.5 in]{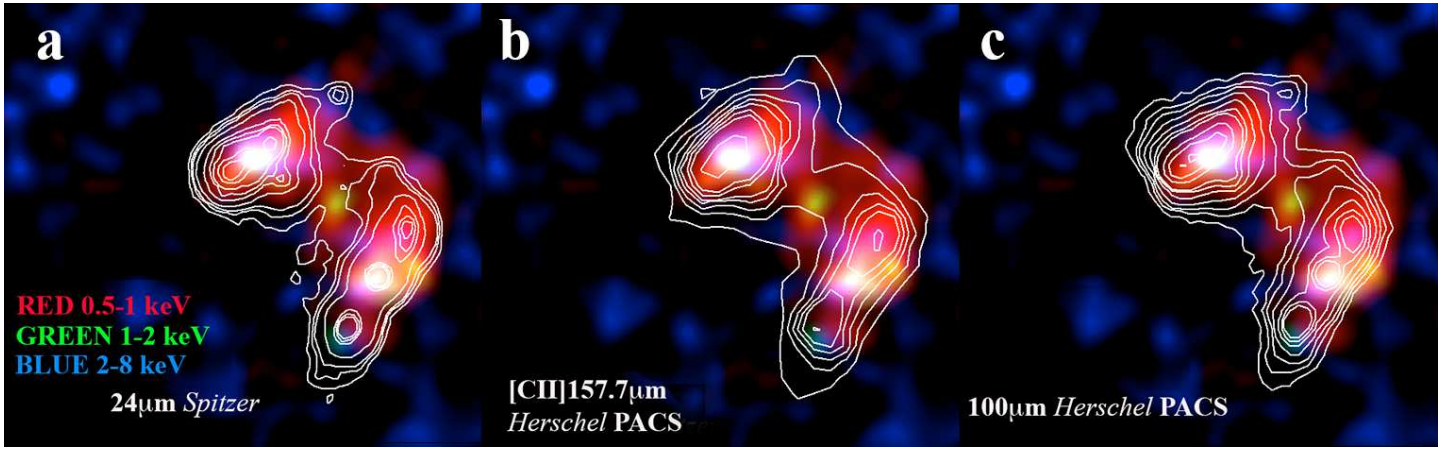}}
\caption{ X-ray color map showing distribution of the soft (0.5--1 keV) X-rays in red, medium hard (1--2 keV) in green and hard (2--8 keV) in blue from {\it Chandra}  as compared with the distribution of (a) 24$\mu$m IR emission from {\it Spitzer}, (b) 157.7$\mu$m integrated [CII] line emission,  and (c)  the 100$\mu$m continuum emission image obtained with the {\it Herschel} observations (Peterson \& Appleton; in preparation). In these images, the 3$\sigma$ compact sources (see text) have been removed before applying the adaptive filtering.}
\end{figure*}

\newpage

\begin{figure*}
\centerline{\includegraphics[width=7in]{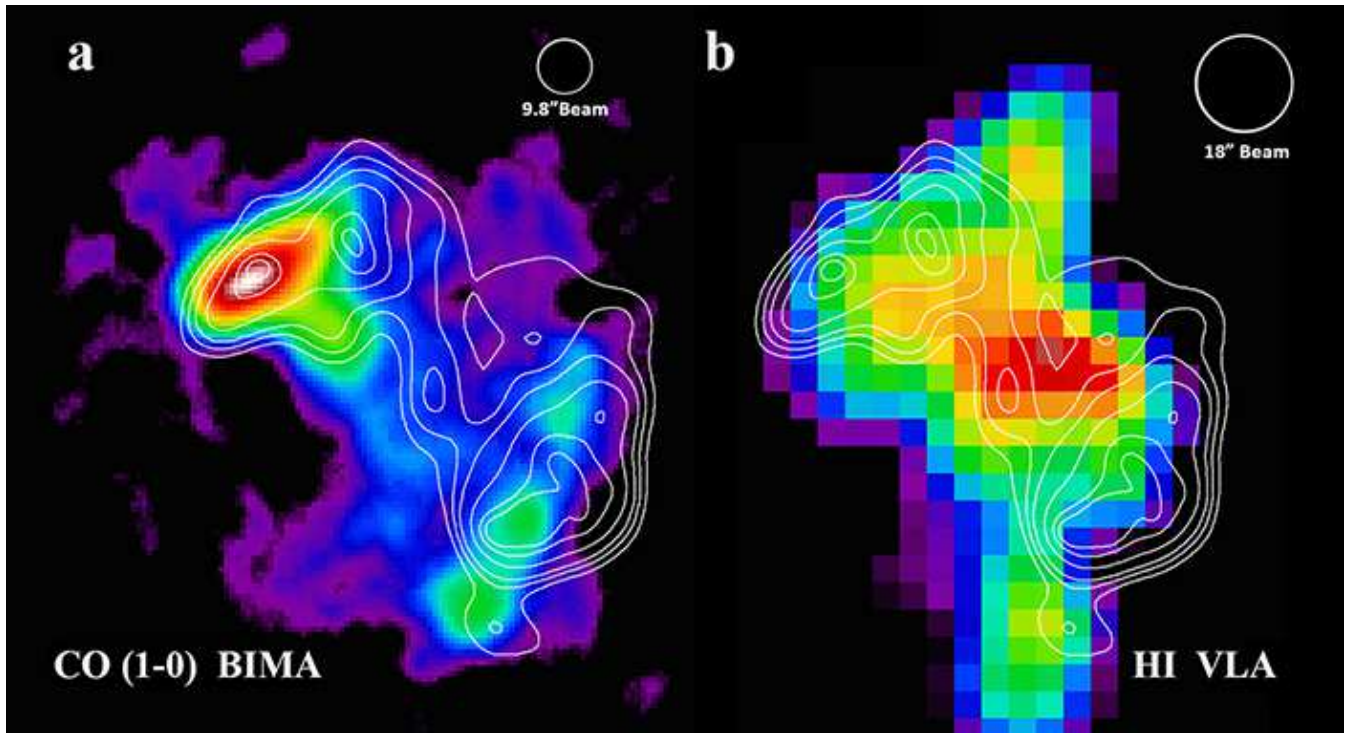}}
\caption{Comparison with cool gas: Soft X-ray (0.5--1 keV) contours (adaptive smoothing with the 3$\sigma$ point sources removed) superimposed on a) the integrated CO (1--0) map from BIMA \citep{gao03a}, and b) the integrated neutral hydrogen map from the VLA \citep{con93}. Contours are in
units of 0.01, 0.0125, 0.015, 0.02, 0.03, 0.04  counts/pixel. One {\it Chandra} pixel has dimension of 0.49 x 0.49 arcsecs$^2$. The CO column densities range from the lowest values of 34 (purple) to 165 Jy beam$^{-1}$ km s$^{-1}$  (white), and for the HI map 4 (purple) to 23 mJy beam$^{-1}$ km s$^{-1}$ with a logarithmic stretch. The BIMA and VLA synthesized (cleaned and restored) beam sizes are 9.8 x 9.8 arcsecs$^2$ and 18 x 18 arcsecs$^2$ respectively (shown as white circles on Figure). }
\end{figure*}

\begin{figure*}
\centerline{\includegraphics[width=4.5in]{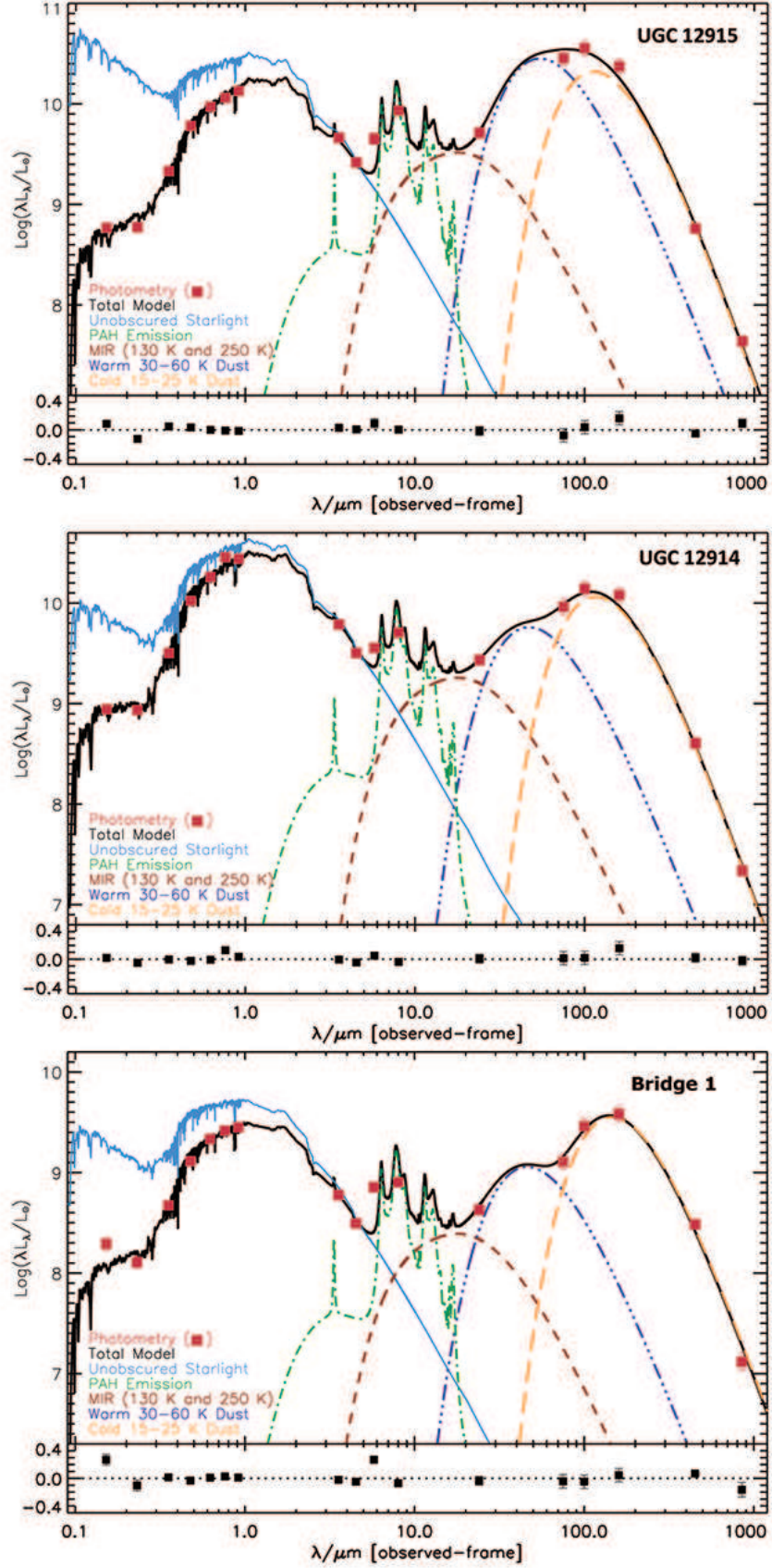}}
\caption{Spectral energy distributions for (a) UGC 12915, (b) UGC 12914, and (c) region Bridge-1 based on data extracted from the regions defined in Figure 2. Data points (red squares) include {\it GALEX}, SDSS, {\it Spitzer} IRAC and MIPS, {\it Herschel} PACS,  and JCMT. The black solid line represents the best-fitting total MAGPHYS model (see text) and the various model components (see key) include unobscured starlight, PAH emission, a Mid-IR, a warm and cold dust component \citep{dac08,dac10}.  }
\label{bridge_spec}
\end{figure*}

\newpage

\begin{figure*}
\centerline{\includegraphics[width=7in]{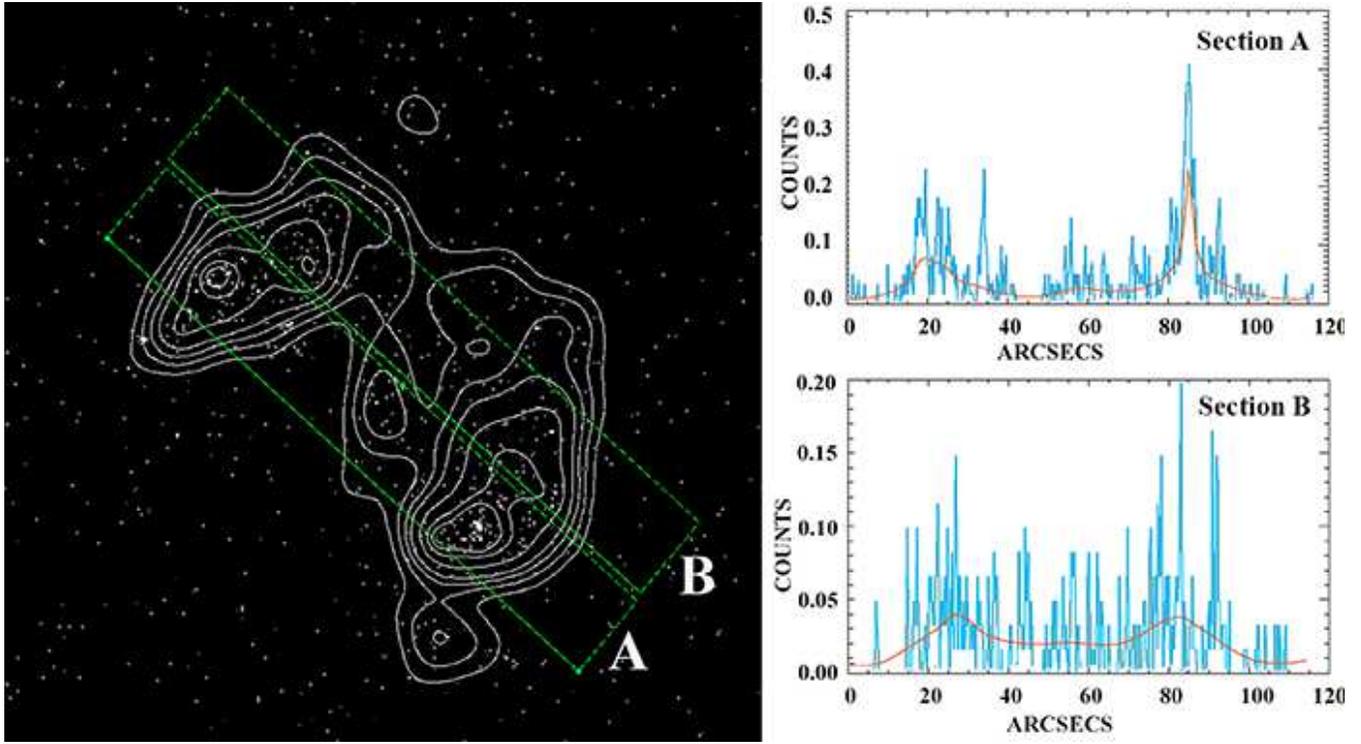}}
\caption{Image cross-sections taken through the Taffy system in soft (0.5--2 keV) X-rays with the 3$\sigma$ compact sources removed. (Left) Unsmoothed counts image showing sections across the galaxy, each section is 120 arcsecs in length integrated over 17 arcsecs in width; contours are of soft-X-rays in the adaptive filter;  (Right)  A plot of the counts in each of the two sections. The blue line are unsmoothed counts, the red line is same data in the  the adaptively filtered map.}
\end{figure*}

\clearpage

\begin{deluxetable*}{llllllll}

\tabletypesize{\small}
\tablecaption{X-ray properties of Selected Regions\label{table1}}
\tablewidth{0pt}
\tablehead{
\colhead{} 	 & \colhead{} 	     & \colhead{} 		 & \colhead{Shape\tablenotemark{a}}       & \colhead{} 		
	& \colhead{} 	    & \colhead{X-ray Emission (0.5-8 keV)} & \colhead{} \\
\cline{3-5}  \cline{6-8}
\colhead{} 	 & \colhead{} 	     & \colhead{} 		 & \colhead{} 		       & \colhead{} 		& \colhead{} 
	& \colhead{Flux\tablenotemark{c,d}} 	& \colhead{Log(L$_X$)}  \\
\colhead{Name} & \colhead{Label\tablenotemark{b}} & \colhead{RA(J2000)} & \colhead{Dec(J2000)} & \colhead{Aperture}	& \colhead{Net Counts}
	&  \colhead{($10^{-15}$\,erg\,s$^{-1}$\,cm$^{-2}$)} & \colhead{(erg\,s$^{-1}$))\tablenotemark{e}} 
}
\startdata
UGC 12914 (diffuse)				&	UGC12914	&	00$^h$01$^m$38.1$^s$	& +23$^d$28$^m$59.0$^s$	& Polygon			& $234\pm24$	& ~~~\,$36.3\pm5.0$		& ~~\,40.22 \\
UGC 12914 (+point sources)$^f$                  &    &     & & & & & ~~\,40.51 \\  
UGC 12915  (diffuse only)			&	UGC12915 	&	00$^h$01$^m$42.1$^s$	& +23$^d$29$^m$47.7$^s$	& Polygon			& $143\pm19$	& ~~~\,$24.9\pm6.0$		& ~~\,40.06 \\
UGC 12915 (+point sources)$^f$				&		&		& 	& 		& 	& 		& ~~\,40.34 \\
Taffy Bridge 				&	Bridge 	&	00$^h$01$^m$40.1$^s$	& +23$^d$29$^m$24.6$^s$	& Polygon			& $101\pm21$	& ~~~\,$11.6\pm4.2$		& ~~\,39.73 \\
Extragalactic HII  (+B1)		&	HII &	00$^h$01$^m$40.9$^s$	& +23$^d$29$^m$37.4$^s$	& Polygon			& $56\pm~\,9$	& ~~~\,$6.4\pm1.1$		&~~\,39.47 \\
\hline
CXOU J000139.0 +232942	&	Bridge-1	&	00$^h$01$^m$39.0$^s$	& +23$^d$29$^m$42.3$^s$	& Polygon			& $85\pm15$	& ~~~\,$9.8\pm$1.7		& ~~\,39.61 \\
CXOU J000141.4 +232903	&	Bridge-2	&	00$^h$01$^m$41.4$^s$	& +23$^d$29$^m$03.5$^s$	& Polygon			& $<$42\tablenotemark{g}~~~~~~\, & ~~~~$<$5.1\tablenotemark{g}~~~~		&$<$39.37\tablenotemark{g}\\
CXOU J000138.3 +232901	&	W$_{nuc}$	&	00$^h$01$^m$38.3$^s$	& +23$^d$29$^m$01.0$^s$	& 2$\farcs$5\,circle	& $90\pm10$ 	& ~~~\,$24.4\pm3.0$		& ~~\,40.06 \\
CXOU J000138.0 +232905	&	W1		&	00$^h$01$^m$38.0$^s$	& +23$^d$29$^m$05.2$^s$	& 2$\farcs$5\,circle 	& $11\pm~\,3$ 	& ~~~\,$2.9\pm0.9$		& ~~\,39.12  \\
CXOU J000139.1 +232909	&	W2		&	00$^h$01$^m$39.1$^s$	& +23$^d$29$^m$09.6$^s$	& 2$\farcs$5\,circle	& $11\pm~\,3$ 	& ~~~\,$3.1\pm1.0$		&~~\,39.12 \\
CXOU J000137.9 +232836	&	W3		&	00$^h$01$^m$37.9$^s$	& +23$^d$28$^m$36.9$^s$	& 2$\farcs$5\,circle 	& $12\pm~\,4$	& ~~~\,$2.9\pm$1.0		& ~~\,39.16  \\
CXOU J000140.9 +232938	&	B1		&	00$^h$01$^m$40.9$^s$	& +23$^d$29$^m$38.1$^s$	& 2$\farcs$5\,circle	& $24\pm~\,5$	& ~~~\,$6.4\pm$1.5		& ~~\,39.47  \\
CXOU J000142.9 +232935	&	E1		&	00$^h$01$^m$42.9$^s$	& +23$^d$29$^m$35.8$^s$	& 2$\farcs$5\,circle	& $20\pm~\,5$	& ~~~\,$5.3\pm1.4$	        & ~~\,39.39  \\
CXOU J000142.3 +232941	&	E2		&	00$^h$01$^m$42.3$^s$	& +23$^d$29$^m$41.1$^s$	& 2$\farcs$5\,circle 	& $20\pm~\,5$	& ~~~\,$5.6\pm1.4$		& ~~\,39.41  \\
CXOU J000141.2 +232949	&	E3		&	00$^h$01$^m$41.2$^s$	& +23$^d$29$^m$49.8$^s$	& 2$\farcs$5\,circle	& $17\pm~\,4$	& ~~~\,$4.5\pm$1.2		& ~~\,39.32  \\
CXOU J000141.6 +232950	&	E4		&	00$^h$01$^m$41.6$^s$	& +23$^d$29$^m$50.4$^s$	& 1$\farcs$5\,circle	& $10\pm~\,3$	& ~~~\,$2.9\pm1.0$		& ~~\,39.12 \\
CXOU J000141.8 +232948	&	E5		&	00$^h$01$^m$41.8$^s$	& +23$^d$29$^m$48.5$^s$	& 1$\farcs$5\,circle 	& $16\pm~\,4$	& ~~~\,$4.5\pm1.2$		& ~~\,39.32  
\enddata
\tablenotetext{a}{Coordinates for the polygon regions are approximate centers. 
Bright point source counts are extracted in a circular aperture \\with a 2$\farcs$5 radius, except where such apertures would overlap significantly. }
\tablenotetext{b}{As labelled in Figure 2a.}
\tablenotetext{c}{ Uncertainties in the point source fluxes are derived from the Poisson uncertainties as well as the uncertainty in the photon index. }
\tablenotetext{d}{Fluxes measured from the fits given in Table \ref{fits} for UGC 12914/5 and Bridge, corrected for foreground absorption The fluxes for Bridge-1, Bridge-2, and HII are calculated using the same model as the full bridge but using the count rate within each region For all of the point sources, we estimate fluxes  from the count rate assuming a power-law model with $\Gamma$=1.7.}
\tablenotetext{e}{Observed luminosities calculated assuming a distance of 62 Mpc.}
\tablenotetext{f}{Total luminosities with inclusion of listed point sources. For UGC 12914 this includes sources W1, W2, W3 and W$_{nut}$,  and for UGC 12915 this includes sources E1 through E5.}
\tablenotetext{g}{3$\sigma$ upper limit}

\end{deluxetable*}

\begin{deluxetable*}{lrrrc}

\tabletypesize{\scriptsize}
\tablecaption{Soft X-ray Emission\label{soft}}
\tablewidth{0pt}
\tablehead{
\colhead{Region} & \colhead{Area} &  \multicolumn{2}{c}{Flux}   & \colhead{Surface} \\
\cline{3-4}
\colhead{}		   & \colhead{} 		 & \colhead{0.5--2\,keV} 		& \colhead{0.3--2\,keV}	& \colhead{Brightness\tablenotemark{a}} \\
\colhead{}		   & \colhead{(as$^2$)} & \multicolumn{2}{c}{($10^{-14}$\,erg\,s$^{-1}$\,cm$^{-2}$)} 		& \colhead{($10^{-18}$\,erg\,s$^{-1}$\,cm$^{-2}$as$^{-2}$)}
}
\startdata
UGC 12914	& 3704	& 	2.9$\pm$0.4	&	3.4$\pm$0.5	& 	\,~9.0	\\
UGC 12915	& 2562	& 	1.6$\pm$0.3	&	1.9$\pm$0.4	& 	\,~7.4	\\
Bridge		& 3771	& 	1.1$\pm$0.3	&	1.2$\pm$0.3	& 	\,~3.1	 \\
Bridge-1		& 1637	& 	0.9$\pm$0.2	&	0.9$\pm$0.2	& 	\,~5.7	 \\
Bridge-2		& 2127	& 	$<$0.5~~~	&	$<$0.5~~~	& 	$<$3~~~		\\
HII			& 356	& 	0.5$\pm$0.2	&	0.6$\pm$0.2	& 	17.1
\enddata
\tablenotetext{a}{In the 0.3-2\,keV band.}
\end{deluxetable*}

\begin{deluxetable*}{lrrlccc}
\tabletypesize{\scriptsize}
\tablecaption{Faint X-ray Compact Sources\label{compact}}
\tablewidth{0pt}
\tablehead{
\colhead{} 	     & \colhead{} 		 & \colhead{Aperture}       & \colhead{} 	& \colhead{Counts Above} & \colhead{Flux (0.5--8keV)\tablenotemark{a}} & \colhead{Log(L$_X$/L$_{\odot}$)}	    \\
\cline{2-4} 
\colhead{} 	     & \colhead{} 		 & \colhead{} 		       & \colhead{} 		& \colhead{Bkgd}  & \colhead{($10^{-15}$\,erg\,s$^{-1}$\,cm$^{-2}$)} &  \\
\colhead{Label} & \colhead{RA(J2000)} & \colhead{Dec(J2000)} & \colhead{Size}	& \colhead{(0.5--8\,keV)}
}
\startdata
B2	&	00$^h$01$^m$39.5$^s$	& +23$^d$29$^m$28.2$^s$	& 2$''$\,circle	& $7.0\pm2.8$	& $1.9\pm0.8 $ & 38.95 \\
B3	&	00$^h$01$^m$40.2$^s$	& +23$^d$29$^m$45.1$^s$	& 2$''$\,circle	& $7.0\pm2.8$  & $1.9\pm0.8 $ & 38.95 \\
B4	&	00$^h$01$^m$40.8$^s$	& +23$^d$29$^m$33.7$^s$	& 2$''$\,circle	& $6.0\pm2.6$	& $1.6\pm0.7 $ & 38.88 \\
B5	&	00$^h$01$^m$39.9$^s$	& +23$^d$29$^m$36.9$^s$	& 2$''$\,circle	& $5.1\pm2.4$ 	 & $1.4\pm0.7 $ & 38.81 \\
E6	&	00$^h$01$^m$40.7$^s$	& +23$^d$30$^m$00.8$^s$	& 2$''$\,circle 	& $6.0\pm2.6$  & $1.6\pm0.7 $ & 38.88 \\
E7	&	00$^h$01$^m$40.8$^s$	& +23$^d$29$^m$54.5$^s$ & 1$\farcs$5\,circle & $5.5\pm2.4$	& $1.5\pm0.7 $ &38.84 \\
E8	&	00$^h$01$^m$40.6$^s$	& +23$^d$29$^m$52.9$^s$ & 1$\farcs$5\,circle & $4.5\pm2.2$	 & $1.2\pm0.6 $ & 38.75 \\
W4	&	00$^h$01$^m$37.5$^s$	& +23$^d$29$^m$37.9$^s$	& 2$''$\,circle 	& $9.1\pm3.2$	 &  $2.5\pm0.9 $ & 39.06 \\
W5	&	00$^h$01$^m$39.3$^s$	& +23$^d$28$^m$32.9$^s$	& 2$''$\,circle	& $7.1\pm2.8$	 & $1.9\pm0.7 $  & 38.95 \\
W6	&	00$^h$01$^m$38.7$^s$	& +23$^d$28$^m$57.2$^s$	& 2$''$\,circle	& $9.1\pm3.2$	& $2.5\pm0.9 $   & 39.06\\
W7	&	00$^h$01$^m$38.4$^s$	& +23$^d$28$^m$51.6$^s$	& 2$''$\,circle	& $5.1\pm2.4$	 & $1.4\pm0.7 $  &  38.81 \\
W8	&	00$^h$01$^m$37.9$^s$	& +23$^d$28$^m$56.1$^s$	& 2$''$\,circle 	& $7.1\pm2.8$	 & $1.9\pm0.7 $ & 38.95 \\
W9	&	00$^h$01$^m$37.6$^s$	& +23$^d$29$^m$01.8$^s$	& 2$''$\,circle	& $6.0\pm2.6$	 &  $1.6\pm0.7 $ & 38.88 \\
W10	&	00$^h$01$^m$37.3$^s$	& +23$^d$29$^m$05.1$^s$	& 2$''$\,circle	& $8.0\pm3.0$	 & $2.2\pm0.8 $ & 39.01 \\
W11	&	00$^h$01$^m$38.1$^s$	& +23$^d$29$^m$10.8$^s$	& 2$''$\,circle 	& $8.0\pm3.0$	  & $2.2\pm0.8 $ & 39.00\\
W12	&	00$^h$01$^m$37.5$^s$	& +23$^d$29$^m$22.9$^s$	& 2$''$\,circle 	& $5.0\pm2.4$	  & $1.4\pm0.7 $  & 38.80 
\enddata
\tablenotetext{a}{Uncertainty in the flux is dominated by the Poisson noise. Fluxes were estimated from the count rate assuming a power law model with $\Gamma$=1.7.}
\end{deluxetable*}

\begin{deluxetable*}{lccccccc}
\tabletypesize{\scriptsize}
\tablecaption{Parameters of X-ray Spectral Fits\label{fits}}
\tablewidth{0pt}
\tablehead{
\colhead{Region} & \colhead{Model\tablenotemark{a}}	& \colhead{N$_{APEC}$\tablenotemark{b}} & \colhead{kT}  
& \colhead{S$_{PL, 1\,keV}$} &  \colhead{$\Gamma$\tablenotemark{c}} & \colhead{N$_{\rm H}$} & \colhead{$\chi^2$/dof} \\
\colhead{} 	& \colhead{} & \colhead{${\rm(10^{-6}\,cm^{-5})}$}& \colhead{(keV)} 
& \colhead{(nJy)} & \colhead{}& \colhead{($10^{22}\,{\rm cm^{2}}$)} & \colhead{} 
}
\startdata

UGC 12914 & APEC+PL 	&  6.1$\pm$3.0		&  0.71$\pm$0.25		 &	3.4$\pm$0.8		& 2.3$\pm$1.5     	& 	...				& ~\,8.4/18	\\
		  & Abs(APEC)	&  28.7$\pm$8.8~	&  0.58$\pm$0.39		 &		...			&	...			& 0.31$\pm$0.09		& 11.9/19		\\
		  & APEC	 	&  8.7$\pm$0.9		&  0.75$\pm$0.06		 &		...			&	...			&	...				& 16.5/20		\\
\tableline	
UGC 12915 & APEC+PL 	& 2.9$\pm$1.2		& 0.72$\pm$0.16      		 &	2.4$\pm$0.7		& 2.0$\pm$0.8  	& 	...				& ~\,7.6/10	\\
		  & Abs(APEC)	& 19.0$\pm$11.9	& 0.61$\pm$0.15  		 &		...			&	...			& 0.39$\pm$0.16		& 10.5/11		\\
		  & APEC	 	& 4.7$\pm$0.8		& 0.78$\pm$0.15 		 &		...			&	...			&	...				& 14.5/12	       \\
\tableline
Taffy Galaxies & APEC+PL & 10.0$\pm$1.7     & 0.73$\pm$0.09                 &     4.3$\pm$1.1              &  1.8$\pm$0.5 	&      ....                     		&~\,21.7/31      \\
(Summed)       & Abs(APEC) & ~397$\pm$383   & 0.26$\pm$0.07                 &          ...                         & ...                        	&   0.58$\pm$0.10             &~\,23.6/32  \\
                         &  APEC       & 13.3$\pm$1.2    & 0.77$\pm$0.05		&     ...				& ...				& ... 					&~\,33.2/33 \\

\tableline
Bridge	  & APEC	 	& 3.8$\pm$0.7		& 0.77$\pm$0.08     		&		...	&	...		&	...			& ~\,4.3/10 	\\
		  & Abs(APEC)	& 10.2$\pm$9.1~	& 0.63$\pm$0.17  		&		...	&	...		& 0.27$\pm$0.18	& \,3.5/9		
\enddata
\tablenotetext{a}{All model have an additional fixed foreground absorption due to Galactic HI of $N_{H}=4.69\times10^{20}\,{\rm cm^{-2}}$. }
\tablenotetext{b}{{\sc apec} normalization is given in units of 10$^{-14} \int n_{e} n_{H} dV / (4 \pi (D_A (1+z))^2)$. In these fits, z=0.0145 and abundance is fixed to solar.}
\tablenotetext{c}{Photon indices, $\Gamma$, are defined in the sense that $P_{E}{\rm (photons~s^{-1}\,keV^{-1})}\propto E^{-\Gamma}$ and relate to the spectral index and flux density with $F_{\nu} \propto \nu^{-\Gamma+1}\propto \nu^{-\alpha}$.}
\end{deluxetable*}

\begin{deluxetable*}{lllccccccc}
\tabletypesize{\scriptsize}
\tablecaption{Derived X-ray Properties \label{model}}
\tablewidth{\linewidth}
\tablehead{
\colhead{Region} & \colhead{Model} & \colhead{kT} & \colhead{$N_{APEC}$} & \colhead{V\tablenotemark{a}} & \colhead{$n_H$\tablenotemark{b}} & \colhead{$M_X$\tablenotemark{c}} &\colhead{$E$\tablenotemark{d}} & \colhead{$P/k$\tablenotemark{e}} & \colhead{T$^f$$_{cool}$}\\
\colhead{} & \colhead{} & \colhead{(keV)} & \colhead{(10$^{-6}$cm$^{-5}$)} & \colhead{(10$^{67}$cm$^{3}$)} & \colhead{(10$^{-3}$cm$^{-3}$)} & \colhead{(10$^8M_{\odot}$)} & \colhead{(10$^{56}$erg)} & \colhead{(10$^4$ K\,cm$^{-3}$) } & \colhead{Gyr}
}
\startdata
Bridge	  & 	APEC		& 0.77		&	3.8		&	5.2		&	1.8	&	0.8		&	1.8		&	2.5    &  1.0    	\\
		  & 	Abs(APEC)  	& 0.63		&	10.2~	&	5.2		&	3.0	&	1.3		&	2.3		&	3.3	& 1.4 	\\
\tableline
UGC 12914 &	APEC+PL		& 0.71		&	6.1		&	8.4		&	1.8	&	1.3		&	2.6		&	2.3	& 1.0$^g$ \\																				
		  &	Abs(APEC)	& 0.58		&	28.7~	&	8.4		&	4.0	&	2.8		&	4.6		&	4.0 & 0.9	\\											
		  &	APEC		& 0.75		&	8.7		&	8.4		&	2.2	&	1.5		&	3.3		&	2.9 & 0.6	\\	
\tableline
UGC 12915 &	APEC+PL		& 0.72		&	2.9		&	4.6		&	1.7	&	0.7		&	1.4		&	2.1	& 1.1$^g$\\																			
		  &	Abs(APEC)	& 0.61		&	19.0~	&	4.6		&	4.2	&	1.7		&	2.9		&	4.6 & 0.8	\\												
		  &	APEC		& 0.78		&	4.7		&	4.6		&	2.2	&	0.8		&	1.9		&	2.9 & 0.5 		
\enddata
\tablenotetext{a}{Assume a Bridge volume of (12.2 kpc$)^3$ and a filling factor of unity; Galaxies assume $\pi$R$^2$h, where R=radius 5.4kpc (4.3kpc) and h=depth 31kpc (27kpc) respectively for UGC 12914 and (UGC12915)} 
\tablenotetext{b}{Derived from the $N_{APEC}$=10$^{-14} \int n_{e} n_{H} dV / (4 \pi (D_A (1+z))^2)$, assuming z=0.0145, D$_A$=61 Mpc, $n_e=n_H$.}
\tablenotetext{c}{$M_X=m_H\times n_H\times V$. We have assumed a filling factor for the X-ray emitting gas of unity so these masses may be overestimated if the gas does not fill the whole assumed volume. }
\tablenotetext{d}{$E=3/2 n_e V k T$}
\tablenotetext{e}{$P/k= 3/2 n_e T$}
\tablenotetext{f}{T$_{cool}$ = E/L$_X$ }
\tablenotetext{g}{Used only the APEC component of the luminosity for the cooling time} 

\end{deluxetable*}

\begin{deluxetable*}{lrrrrr}
\tabletypesize{\scriptsize}
\tablecaption{Infrared properties and Star Formation Rates Derived from SED Fitting\label{ir}}
\tablewidth{0pt}
\tablehead{
\colhead {Region\tablenotemark{a}}        & \colhead{ log(L$_{TIR}$/L$_{\odot}$)\tablenotemark{b}}     & \colhead {log(L$_{FIR}$/L$_{\odot}$)\tablenotemark{c}}    & \colhead {log(M$_{dust}$)/M$_{\odot}$)}     & 
\colhead{log(M$_{stars}$/M$_{\odot}$)} & \colhead{SFR (M$_{\odot}$\,yr$^{-1}$)} 
}
\startdata
\vspace{1mm}
UGC 12914   & 10.43$^{+0.03}_{-0.01}$   & 10.01$^{+0.03}_{-0.01}$   & 7.42$\pm$0.06  & 10.87$^{+0.12}_{-0.01}$	& 1.05$^{+0.08}_{-0.13}$  \\
\vspace{1mm}
UGC 12915   & 10.81$^{+0.02}_{-0.07}$   & 10.54$^{+0.02}_{-0.07}$   & 7.56$\pm$0.06 &  10.62$^{+0.07}_{-0.21}$		& 2.55$^{+0.24}_{-1.10}$  \\
\vspace{1mm}
Bridge-1	     &   9.79$^{+0.02}_{-0.04}$   &   9.28$^{+0.02}_{-0.04}$   & 7.50$\pm$0.06  	& 9.87$^{+0.09}_{-0.06}$	& 0.26$^{+0.02}_{-0.04}$ \\
\vspace{1mm}
Bridge-2         &   9.66$^{+0.02}_{-0.03}$  &   9.14$^{+0.02}_{-0.03}$    & 7.59$\pm$0.04  	& 9.77$^{+0.07}_{-0.15}$	& 0.19$^{+0.02}_{-0.01}$ \\
\vspace{1mm}
HII     	    &    9.64$^{+0.02}_{-0.01}$   &   9.22$^{+0.02}_{-0.01}$   & 7.30$^{+0.09}_{-0.06}$ & 9.04$^{+0.02}_{-0.05}$ & 0.24$^{+0.03}_{-0.06}$ 
\enddata
\tablenotetext{a}{Regions defined from Figure 2.}
\tablenotetext{b}{TIR is defined as $3-1000\mu{\rm m}$.}
\tablenotetext{c}{FIR is defined as $42-122\mu{\rm m}$.}

\end{deluxetable*}


\end{document}